\documentclass[pra,twocolumn,superscriptaddress,nofootinbib]{revtex4}

\usepackage{amsmath,amsfonts,amssymb,amstext}
\usepackage[dvipdfm]{graphicx}
\usepackage[colorlinks,linkcolor=blue]{hyperref}

\newcommand{\cO}[1]{{\mathcal O}(#1)}

\begin{document}
\title{Phase transition of computational power in the resource states \\
for one-way quantum computation}
\date{September 12, 2007}

\author{Daniel E. Browne}
\affiliation{Department of Materials, University of Oxford, Parks Road, Oxford, OX1 3PH, UK}
\affiliation{Department of Physics and Astronomy, University College London, Gower Street, London WC1E 6BT, UK}

\author{Matthew B. Elliott}
\author{Steven T. Flammia}
\author{Seth T. Merkel}
\affiliation{Department of Physics and Astronomy, University of New Mexico, Albuquerque, NM, 87131, USA}

\author{Akimasa Miyake}
\affiliation{\mbox{Institute for Theoretical Physics, University of Innsbruck,
Technikerstra{\ss}e 25, A-6020 Innsbruck, Austria}}
\affiliation{\mbox{Institute for Quantum Optics and Quantum Information,
Austrian Academy of Sciences, Innsbruck, Austria}}

\author{Anthony J. Short}
\affiliation{Department of Applied Mathematics and Theoretical Physics, University
of Cambridge, Wilberforce Road, Cambridge CB3 0WA, UK}

\begin{abstract}
We study how heralded qubit losses during the preparation of a two-dimensional cluster state, a universal resource state for one-way quantum computation, affect its computational power. Above the percolation threshold we present a polynomial-time algorithm that concentrates a universal cluster state, using  resources that scale optimally in the size of the original lattice. On the other hand, below the percolation threshold, we show that single qubit measurements on the faulty lattice can be efficiently simulated classically. We observe a phase transition at the threshold when the amount of entanglement in the faulty lattice directly relevant to the computational power changes exponentially.
\end{abstract}

\maketitle

\section{Introduction}

The one-way model for quantum computation~\cite{Raussendorf2001}, or measurement-based quantum computation (MBQC), is an alternative to the standard circuit model~\cite{Nielsen2000}, providing another possible route towards practical quantum information processing. In this scheme, quantum computation proceeds by adaptive single-qubit measurement on an entangled universal resource state.  Particular physical implementations, however, may necessitate modifying the original scheme, which used a cluster state \cite{Briegel2001} on the $2$-dimensional square lattice as its universal resource. Other resource states that have been identified as universal under this measurement-based scheme include the cluster state corresponding to a $2$-dimensional hexagonal lattice~\cite{Van-den-Nest2006,Van-den-Nest2007a}, and, in addition, other extended measurement-based schemes have been proposed~\cite{Gross2007}. In this paper, rather than searching for classes of universal states, we consider the impact of possible experimental imperfections on the resource states to see if universal MBQC can be achieved. We also show that our model exhibits a phase transition in the computational power of the resource state, in the sense that its potential for universal quantum computation is lost suddenly at a particular critical degree of imperfection. The notion of universality we use throughout is that of \emph{efficient} universality (in the sense of  Ref.~\cite{Van-den-Nest2007a}) i.e. that any quantum computation can be implemented with only a polynomial overhead in spatial resources and time \footnote{More formally, we require that the output states of any family of poly($n$)-sized quantum circuits on $n$ qubits can be generated using a resource state of size poly($n$), and in time poly($n$) (both in terms of quantum measurements and classical side-processing).}, as efficiency is a crucial aspect of computational power.

Typically, one imagines a cluster state as being prepared by placing qubits in an $L \times L$ lattice and then entangling the qubits with an Ising type interaction between nearest neighbours.  The model we consider in this paper is one in which there is only a finite probability $p$ that there was initially an atom in any given lattice site.  The value of $p$ is independent and identical for each lattice site.  We could imagine this situation arising when we fill an optical lattice with atoms from a Mott-insulator with some sub-unital filling factor \cite{DePue1999,Jaksch1999,Mandel2003}.  We assume there is only at most one atom per site, and furthermore that the atom positions are heralded.  The resulting state looks like a regular lattice graph state, but with some vertices and their incident edges removed.

Percolation is the widely-known phenomenon that in lattices with imperfect filling there exists a unique giant connected cluster of size $\cO N$ (where $N = L^2$), when the filling factor lies above a certain percolation threshold (in the so-called supercritical phase). One might naively think it is obvious that above this threshold the corresponding state in our model would be  universal as a resource of MBQC.  However, a connected graph state of size $\cO{N}$ is generally not sufficient to be a universal resource.  For example, a 1D chain (i.e., 1D cluster state) of $\cO{N}$ is not universal, nor is any $\cO{N}$ tree structure nor even the complete graph on $N$ vertices \cite{Van-den-Nest2006}.  In this paper, we show that the faulty lattice is indeed universal by showing explicitly how to convert the supercritical resource state into the regular 2D cluster state with constant overhead (dependent only on $p$) and polynomial-time classical processing.

Finally, we interpret the transition to universality at the percolation threshold as a phase transition, with the entanglement of the faulty lattice as an order parameter. Specifically, we use an entanglement measure called the \textit{entanglement width\/}~\cite{Van-den-Nest2006,Van-den-Nest2007a} which is well suited as an entanglement measure for states being used in MBQC. We show that the scaling exponent of the entanglement width of the faulty lattice goes through a discontinuity at the critical probability $p_c$ in the large $N$ limit.

It should be noted that in this paper all statistical statements, such as discussion of phase transitions, should be interpreted in the following way.  We consider properties that hold true almost surely (more precisely, with probability exponentially close to one in the thermodynamic limit), for all the possible realizations of the configuration of holes with a given $p$. This paper does not directly consider mixed states, although clearly the mixed state corresponding to a mixture of states weighted according to their probability for a given $p$, would reduce to the same ensemble of pure states under a non-demolition site-occupation number measurement.

We briefly mention the literature which utilized the idea of percolation theory for the quantum information processing.
A simple modification of the situation above is to consider the effect when all lattice sites are filled with an atom but the entangling gate fails with some probability, leading to  a graph state with missing edges instead of vertices.  This is the model considered in Ref.~\cite{Kieling2006}, motivated by linear optical implementations of quantum computation.  In that paper the authors consider an initial faulty resource of size $L^{1+\epsilon} \times L^{1+\epsilon} \times L^{\epsilon}$, a $3$-dimensional resource, and show that above the edge percolation threshold for a $3$D lattice, it is possible to concentrate an $\cO{L} \times \cO{L}$ hexagonal cluster state.  The approach used in Refs.~\cite{Kieling2006,Gross2007a} is derived from renormalization theory and is applicable to a wide variety of lattice geometries, and may also be adapted to work with vertex defects, rather than edge defects. We note that in the context of quantum networks, on the other hand, classical percolation methods were not found to be optimal \cite{Acin2007}.


\section{Preliminaries}

Because this paper relies on results from rather distinct disciplines, this section provides the necessary background for readers who may be unfamiliar with one or more topics.

\subsection{Graph theory}

\begin{figure}[t]
\begin{center}
\includegraphics[width=8cm,clip]{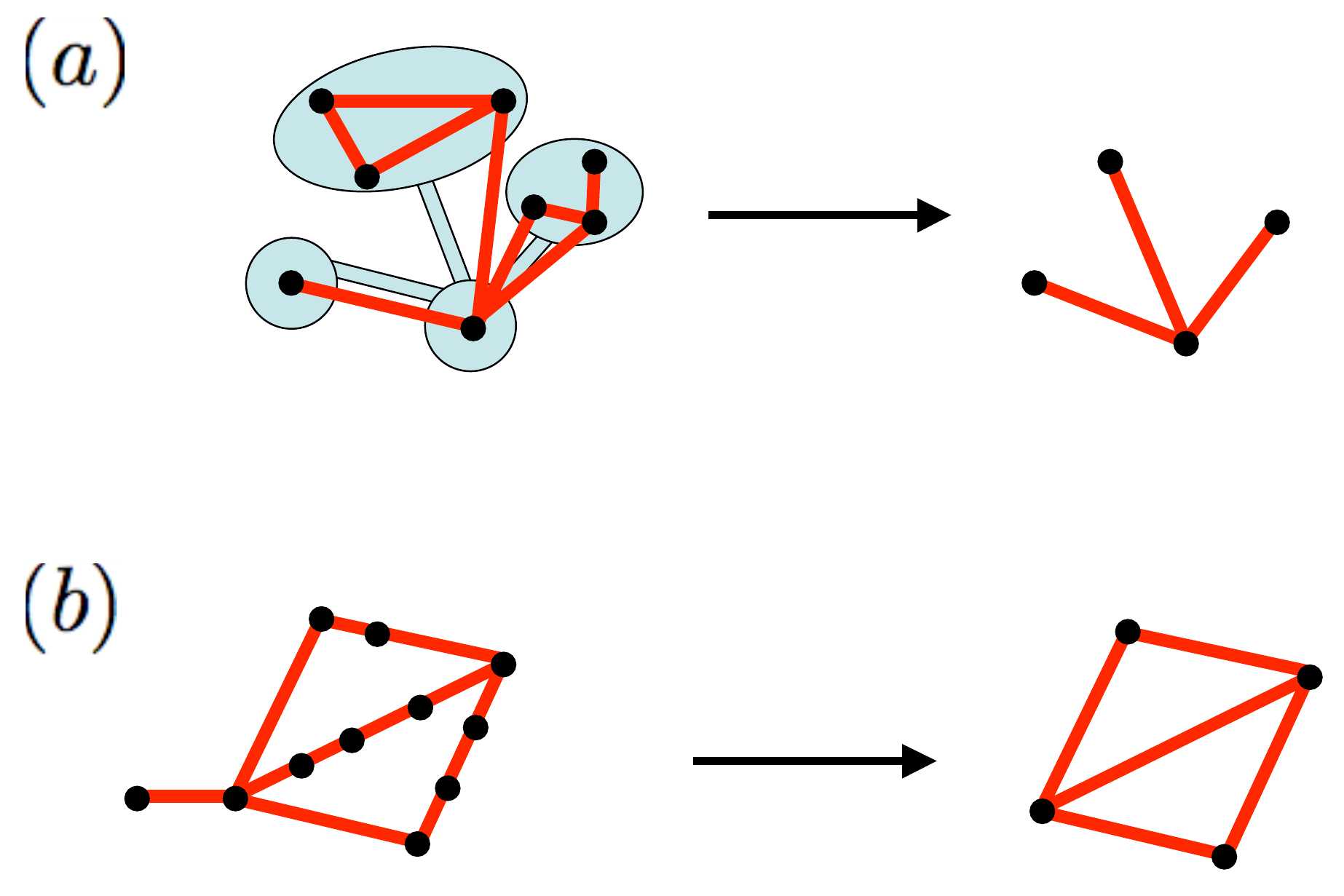}
\caption{(a) Example of a graph minor; (b) example of a topological minor.}
\label{F:minor}
\end{center}
\end{figure}

See Ref.~\cite{Diestel2000}. A \textit{graph} $G$ is a set of points, called \textit{vertices}, along with a set of pairs of vertices which are called \textit{edges}. Vertices are usually labeled with numbers, such as vertex $j$, and edges are usually labeled by the vertices they connect, such as edge $\{j,k\}$ that connects vertices $j$ and $k$. The vertices $j$ and $k$ are sometimes said to be the \textit{endpoints} of the edge, or be \textit{connected} by the edge, and also sometimes the edge is said to be \textit{incident} on vertices $j$ and $k$. A graph is represented pictorially with dots representing vertices and lines connecting the dots representing edges. Two vertices connected by an edge are said to be \textit{neighbors}, and the number of neighbors of a vertex is dubbed its \textit{degree}. The \textit{neighborhood} of vertex $j$, denoted as ${\mathcal N}(j)$, is the set of vertices that are neighbors of vertex $j$. The neighborhood of a set of vertices is the union of the neighborhoods of the vertices in the set (excluding the vertices themselves). A \textit{path} in $G$ is an ordered set of vertices and their connecting edges, $\{v_1,v_2,\ldots,v_k\}$ and $\{\{v_1,v_2\},\{v_2,v_3\},\ldots,\{v_{k-1},v_k\}\}$, and each vertex and edge appear exactly once.  Thus, paths have no ``dangling'' edges and cannot self-intersect.  The first and last vertices in the path are the \textit{starting} and \textit{ending} vertices, respectively.

There are two more important concepts we need to define: graph minors and topological minors.  These are illustrated in Figure~\ref{F:minor}. A {\it graph minor\/}, or simply minor, is obtained by first partitioning the vertices into disjoint connected sets, and then considering the graph between these sets, while disregarding any self-loops or parallel edges; see Figure~\ref{F:minor}(a).  A {\it topological minor\/} is a special case of a graph minor where the resulting minor has the same topology as the original graph.  More precisely, if we view the original graph as a topological space, a topological minor is homeomorphic to the original graph.  This is illustrated in Figure~\ref{F:minor}(b).  For a rigorous definition of these concepts, see \cite{Diestel2000}.

\begin{figure}[t]
\begin{center}
\includegraphics[width=8cm,clip]{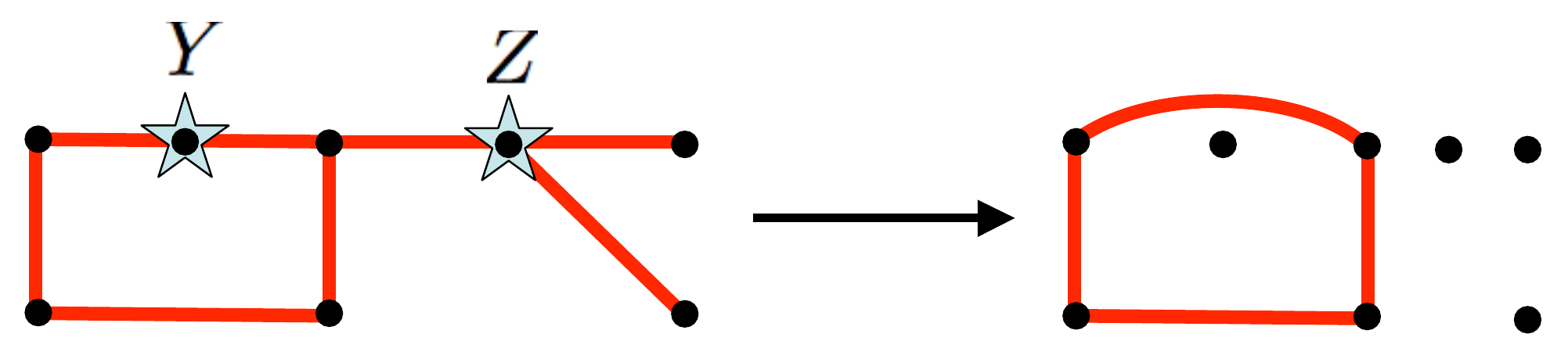}
\caption{Effects of measuring $Y$ and $Z$ on a graph state. The graph on the right is local unitary equivalent to the post-measurement state.}
\label{F:YZmeasure}
\end{center}
\end{figure}

\subsection{Graph states, cluster states, and Pauli measurements}

The universal resource states we address are called \textit{graph states} \cite{Hein2005} and are defined in terms of a graph $G$ as follows. Let $X_j$, $Y_j$, and $Z_j$ denote the Pauli operators on the $j$-th qubit. For each vertex $j$ of the graph $G$, prepare a qubit initially in the $+1$ eigenstate of $X_j$. Then apply the controlled-Z operator between each pair of qubits who are neighbors in $G$. For a connected graph, what results is an entangled pure quantum state that is well represented by the underlying graph $G$.
The class of graph states resulting from $2$D regular lattices are called \textit{cluster states\/}. The two lattices that we use are the square lattice and the hexagonal lattice. Both have been shown to be universal for one-way quantum computation \cite{Raussendorf2001,Van-den-Nest2006}.
\begin{figure*}[t!]
\begin{center}
\begin{tabular}{ccc}
\includegraphics[width=5.9cm,clip]{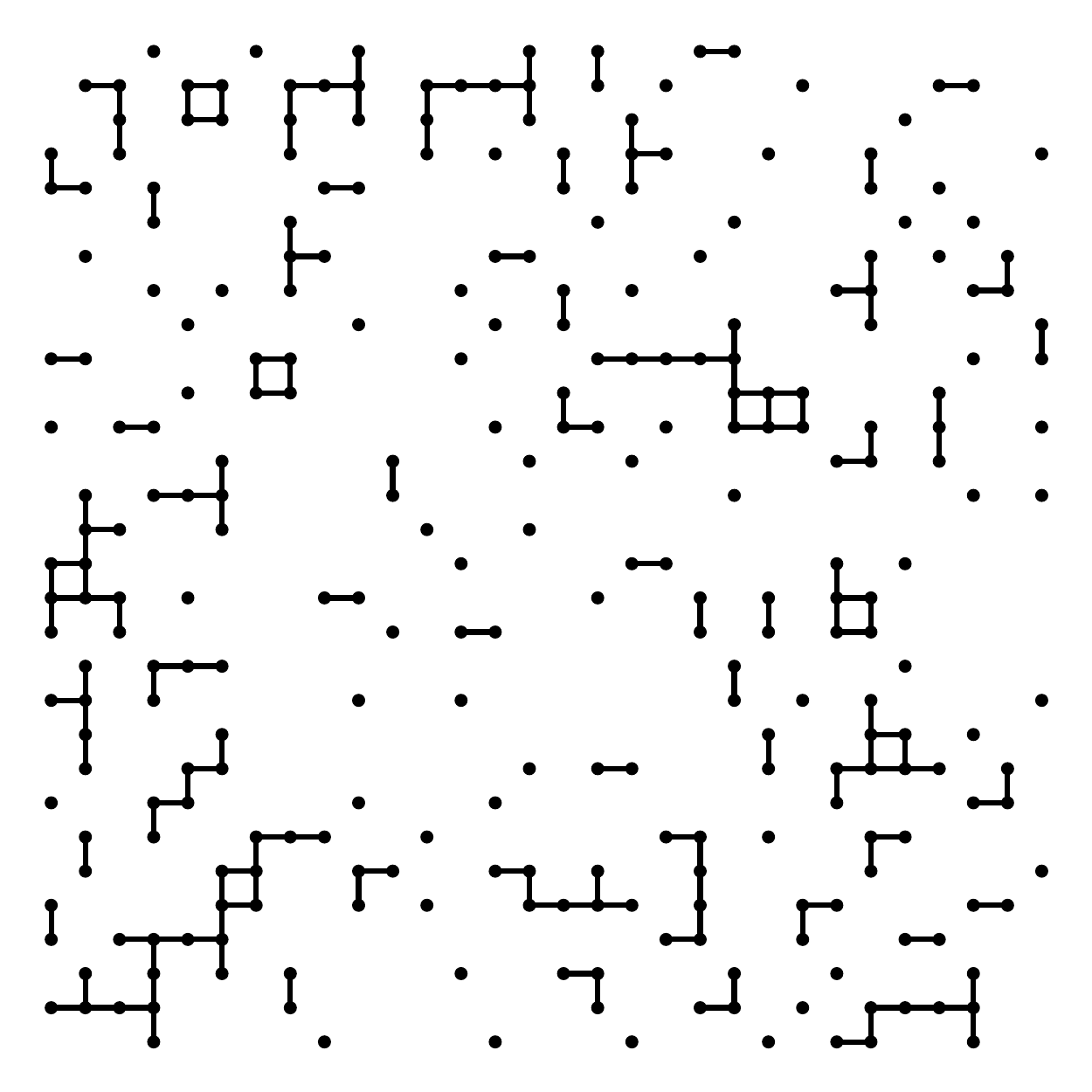} &
\includegraphics[width=5.9cm,clip]{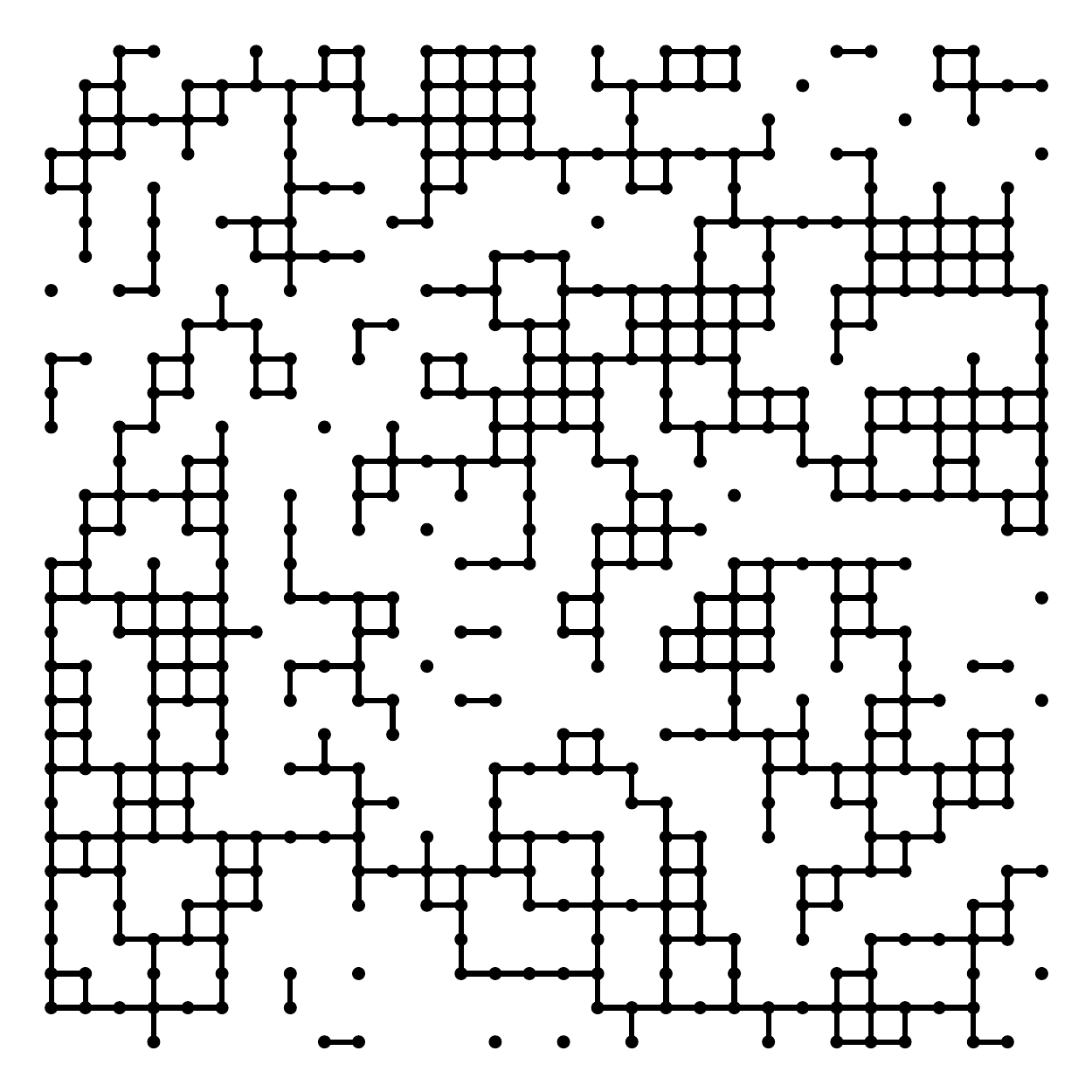} &
\includegraphics[width=5.9cm,clip]{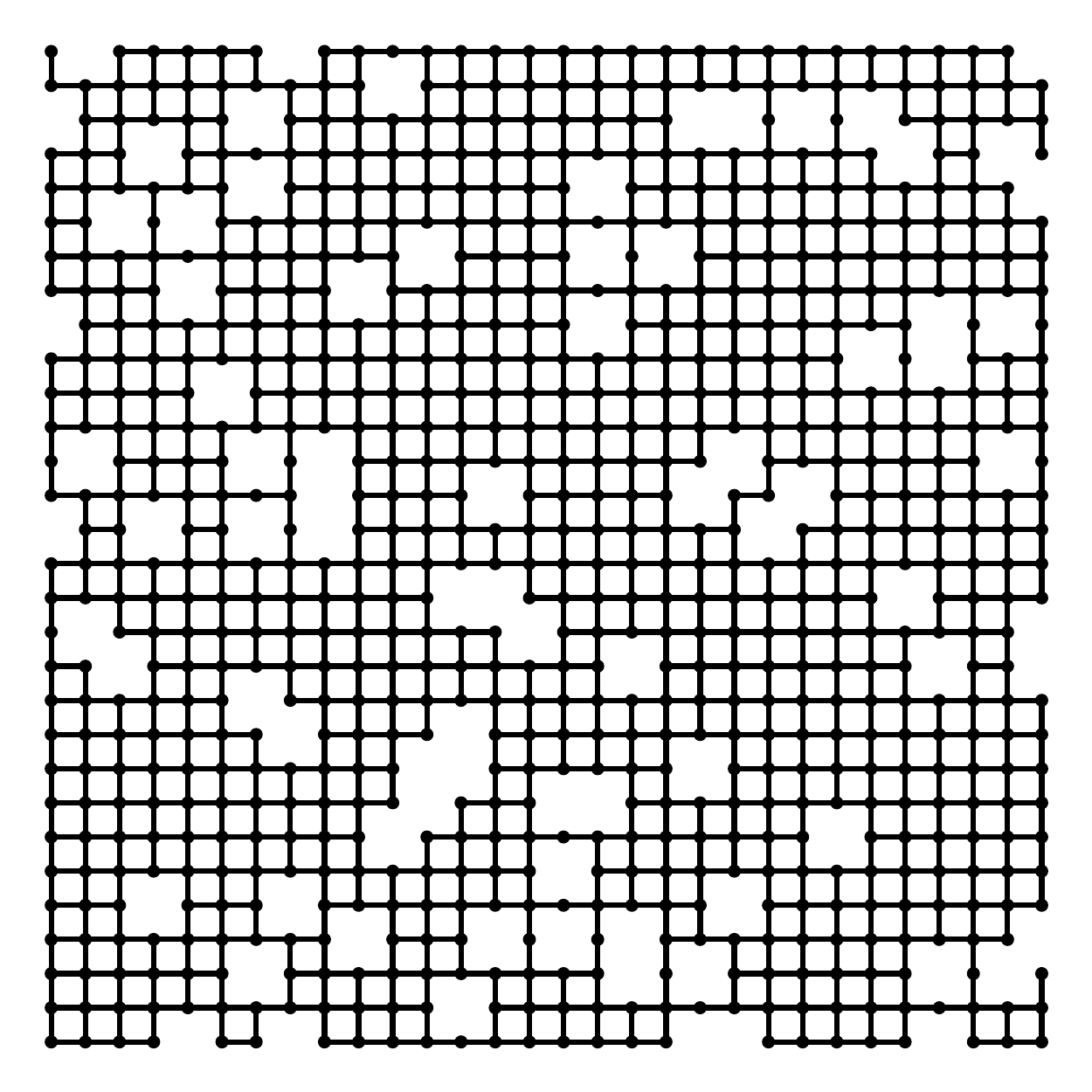} \\
(a) & (b) & (c)
\end{tabular}
\caption{Example of site percolation on a $30 \times 30$ lattice for site probabilities in each of the three phases: (a) is subcritical with $p=.3$, (b) is approximately critical with $p=.592746 \approx p_c$, and (c) is supercritical with $p=.9$.  The occupied sites are colored black.  The same random seed was used for each phase to facilitate comparison.  For (a), which is deep in the subcritical phase, there is no crossing.  Similarly, (c) is deep in the supercritical phase and there are many crossings.  In~(b), we are at the percolation threshold, and there is no vertical crossing, but there are two horizontal crossings.}
\label{F:perc-example}
\end{center}
\end{figure*}

The effects of Pauli measurements on graph states are given in Ref.~\cite{Hein2004}. For simplicity, we describe here the effect of Pauli measurements on graph states for the two cases we need. If a measurement of $Z_j$ is made on a graph state represented by a graph $G$, then the resulting state is local unitarily equivalent to a graph state with graph $G'$, which results from $G$ by removing all edges incident on vertex $j$. We refer to this process as \textit{disconnecting} qubit $j$. On the other hand, if a measurement of $Y_j$ is made on the state represented by $G$, and if vertex $j$ has degree $2$, the resulting graph $G'$ is obtained from $G$ by connecting the two neighbors of vertex $j$ before disconnecting vertex $j$ from the graph. We refer to this process as \textit{contracting} the line containing qubit $j$. Note that this process preserves the topology of the initial graph.  These two processes are illustrated in Figure~\ref{F:YZmeasure}. The post-measurement state depends on the measurement outcome, but the two possible states are equivalent to graph states up to known local Clifford operations. These do not need to be actively corrected. They can be recorded and incorporated into the choice of bases for later measurements.

In this paper, we need only to disconnect qubits and contract lines in order to concentrate a perfect lattice from a faulty one. While it could be that other measurement configurations might improve the efficiency of this process, they could only do so by a constant factor since we already achieve the optimal scaling. Moreover, other measurement configurations tend to result in very complicated graph transformations, so this possibility seems unlikely.

\subsection{Percolation theory}\label{S:perctheory}

{\it Percolation theory\/} is the study of simple lattice models with defects.  In {\it site\/} percolation, we consider a regular lattice, say a square lattice, having a site {\it occupied\/} with probability $p$, and {\it unoccupied\/} with probability $1-p$.  We then imagine that nearest neighbors on the lattice are connected by edges if both neighbors are occupied.  A particular instantiation of a $30 \times 30$ square lattice for various values of $p$ is shown in Figure~\ref{F:perc-example}.

We are interested in statistical properties of this ensemble for lattices of finite extent, say $L \times L$, but in the limit of large $L$.  One such property is the existence of a {\it crossing\/}.  An {\it H-crossing} (short for {\it horizontal crossing\/}) is a path through neighboring occupied sites from the left-hand boundary to the right-hand boundary.  Similarly, a {\it V-crossing\/} traverses the graph from bottom to top.

Does there exist a critical value $p = p_c$ above which we are guaranteed to find a crossing in the large $L$ limit?  The answer is yes \cite{Grimmett1989}, and this is the central result in percolation theory.  The value $p_c$, called the {\it threshold\/}, depends on the lattice geometry.  Since we only concern ourselves in this paper with site percolation on the square lattice, it is appropriate to specialize to this case; there the threshold is $p_c \approx .592746$.  There are then three distinct phases in the model: the {\it subcritical\/} phase where $p< p_c$, the {\it critical\/} phase where $p= p_c$, and the {\it supercritical\/} phase where $p> p_c$.  These three phases are illustrated in Figure~\ref{F:perc-example}.

In the supercritical phase, there exists {\em at least\/} one H-crossing (or V-crossing), but how many are there?  As discussed in the Appendix, there are $\cO L$ vertex-disjoint crossings almost surely.

\begin{figure*}
\begin{center}
\begin{tabular}{ccc}
\includegraphics[width=7.1cm,clip]{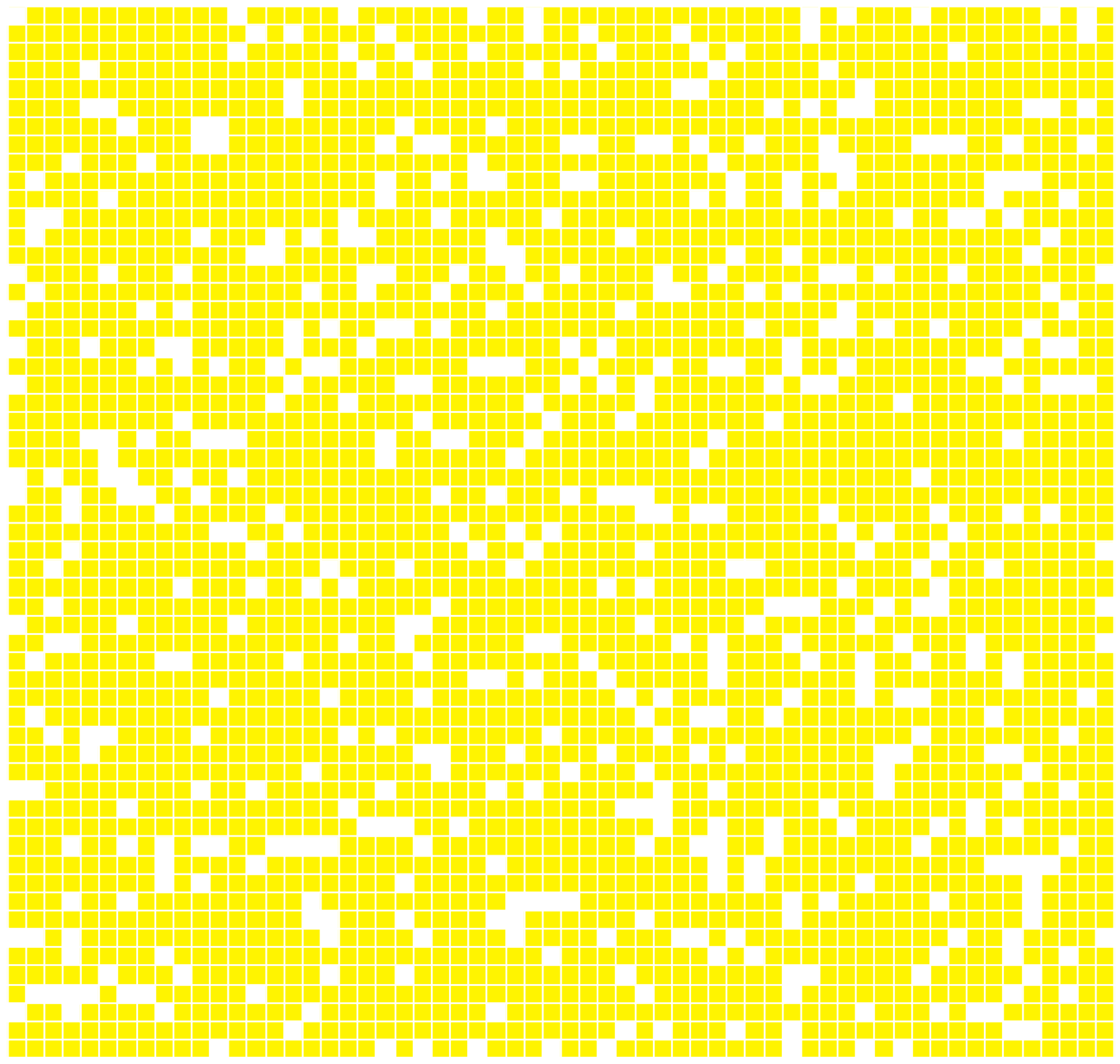} &$\quad$&
\includegraphics[width=7.1cm,clip]{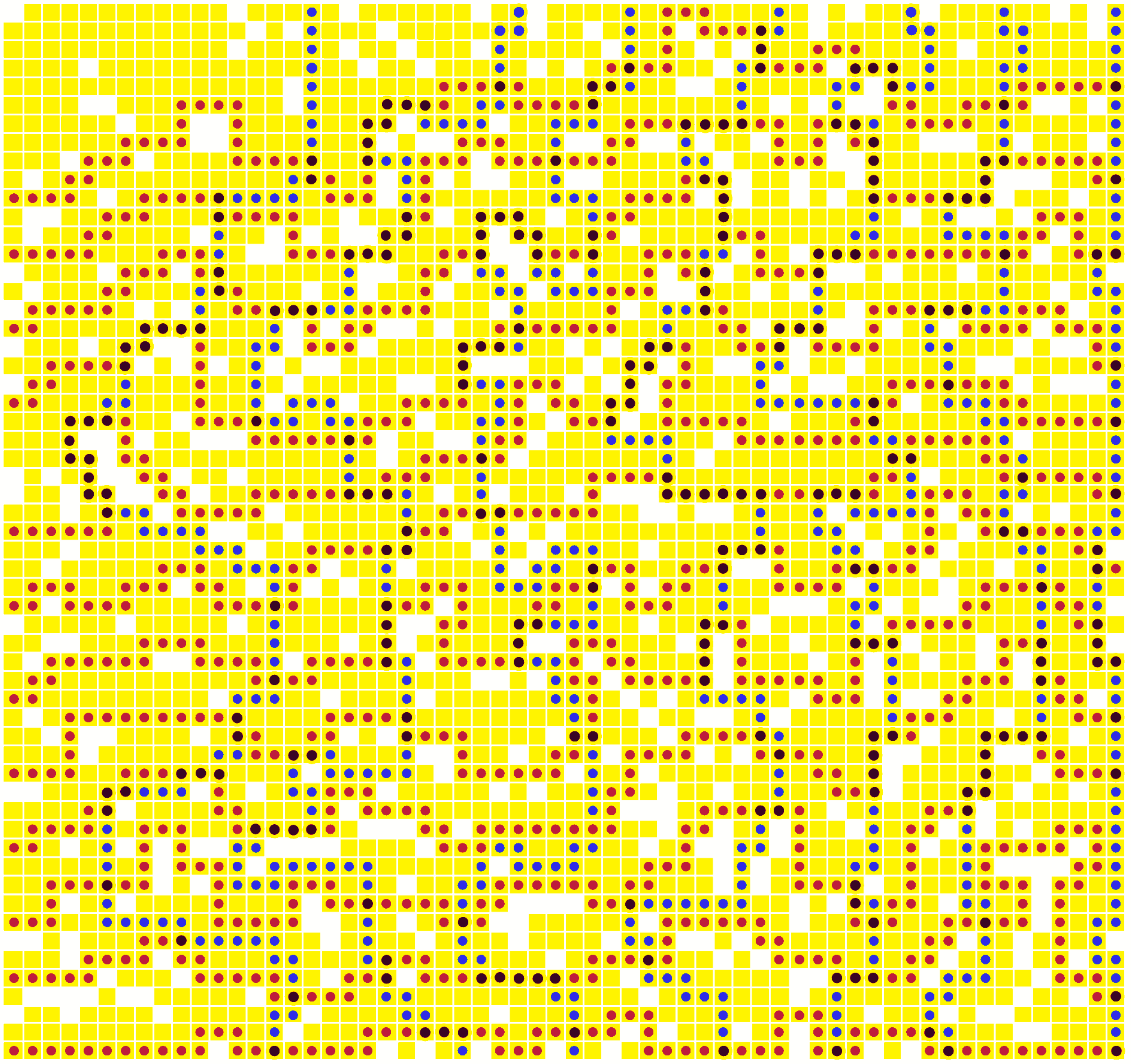} \\
(a) initial faulty square lattice &$\quad$&
(b) path identification (C.1) \\
\includegraphics[width=7.1cm,clip]{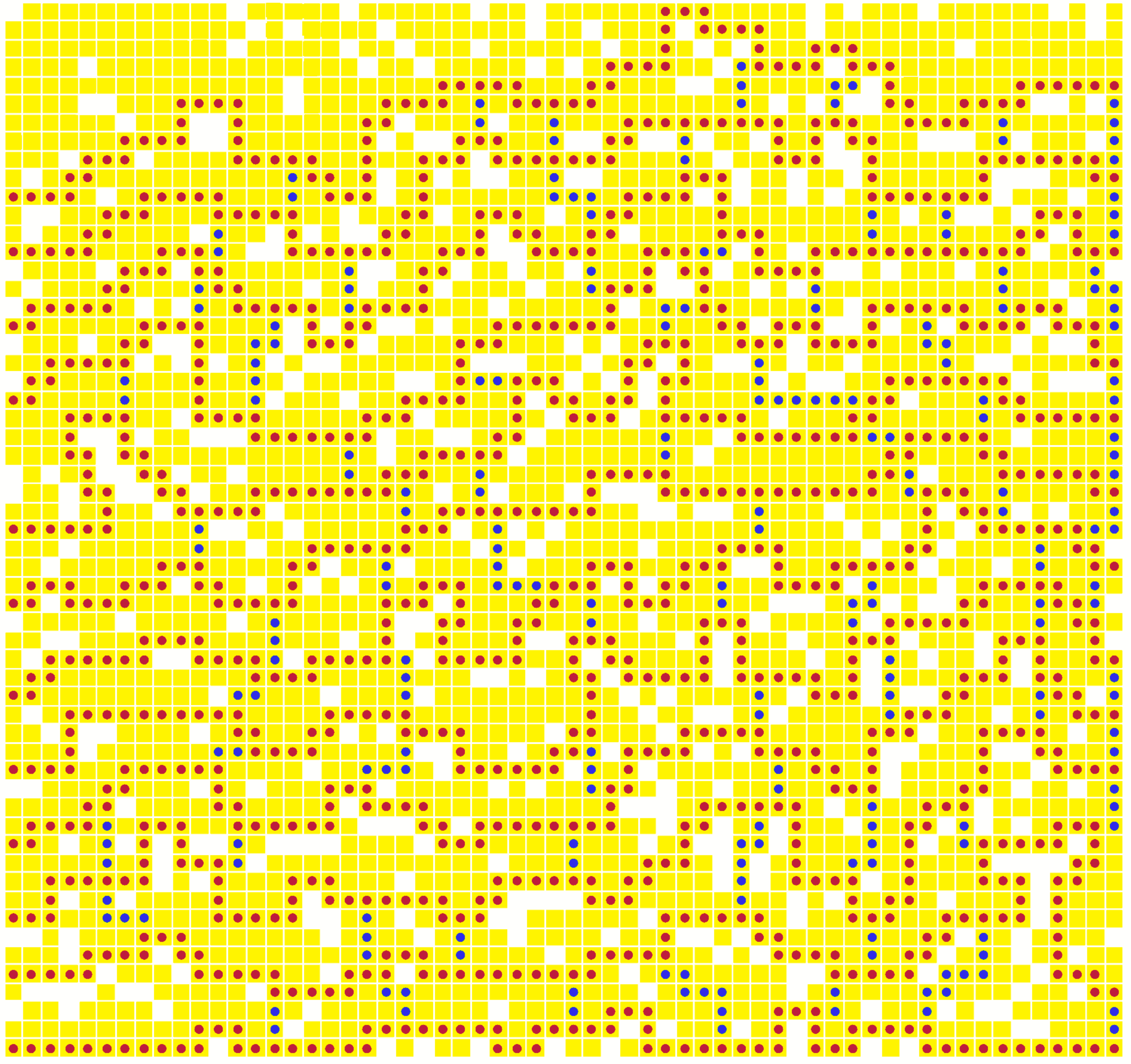} &$\quad$&
\includegraphics[width=7.1cm,clip]{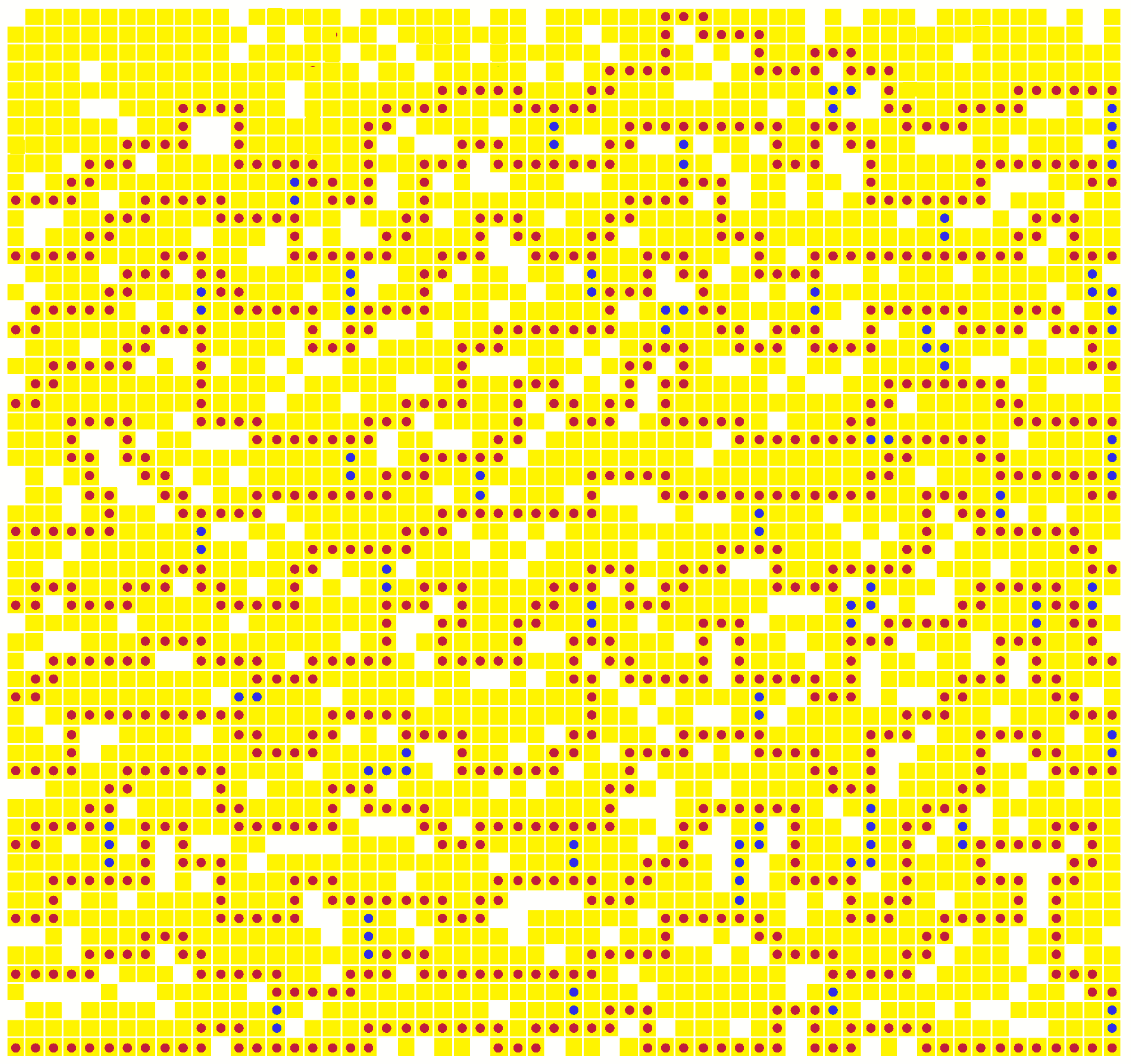} \\
(c) bridge decomposition (C.2) &$\quad$&
(d) alternating bridge decomposition (C.2) \\
\includegraphics[width=7.1cm,clip]{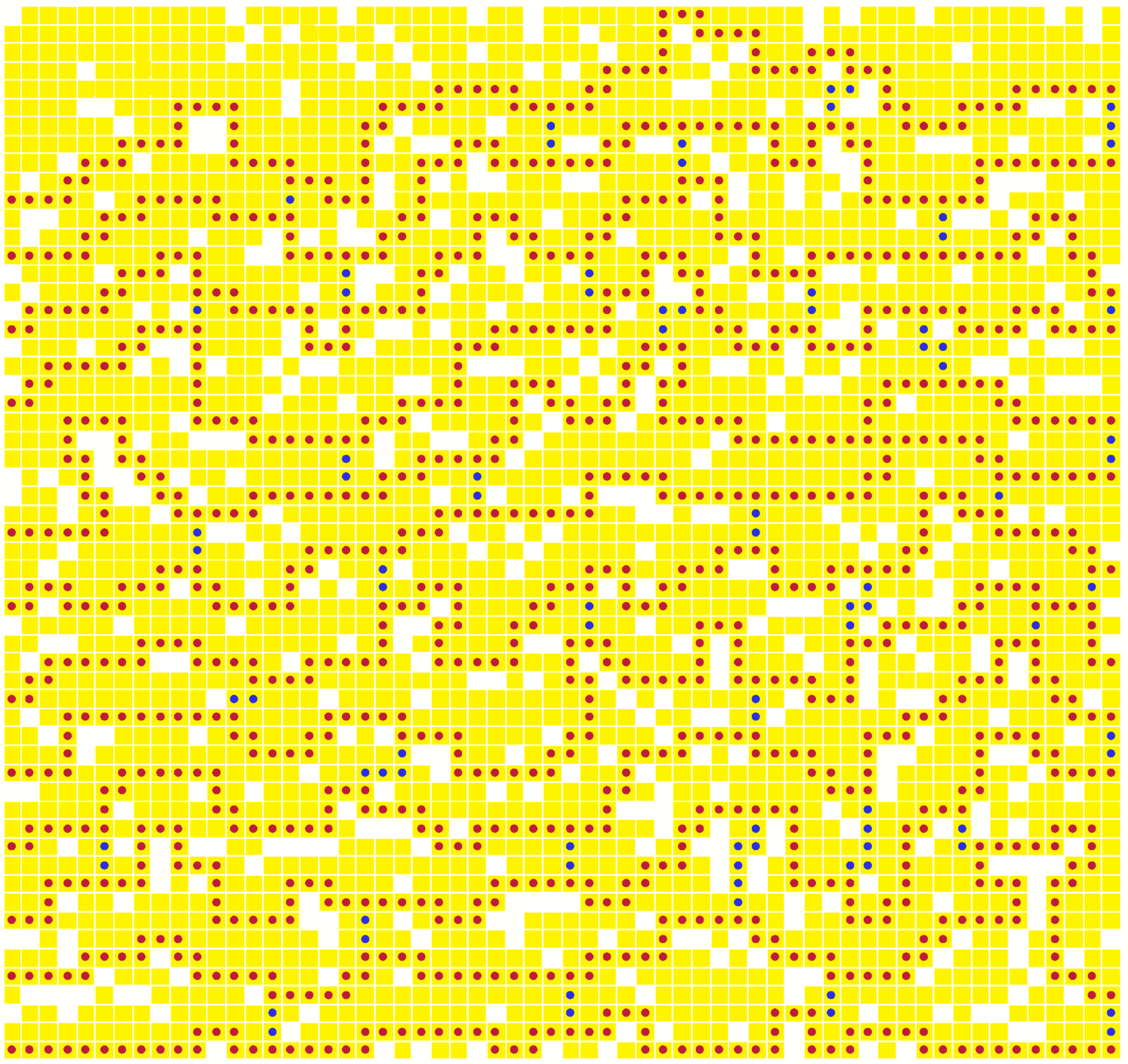} &$\quad$&
\includegraphics[width=7.1cm,clip]{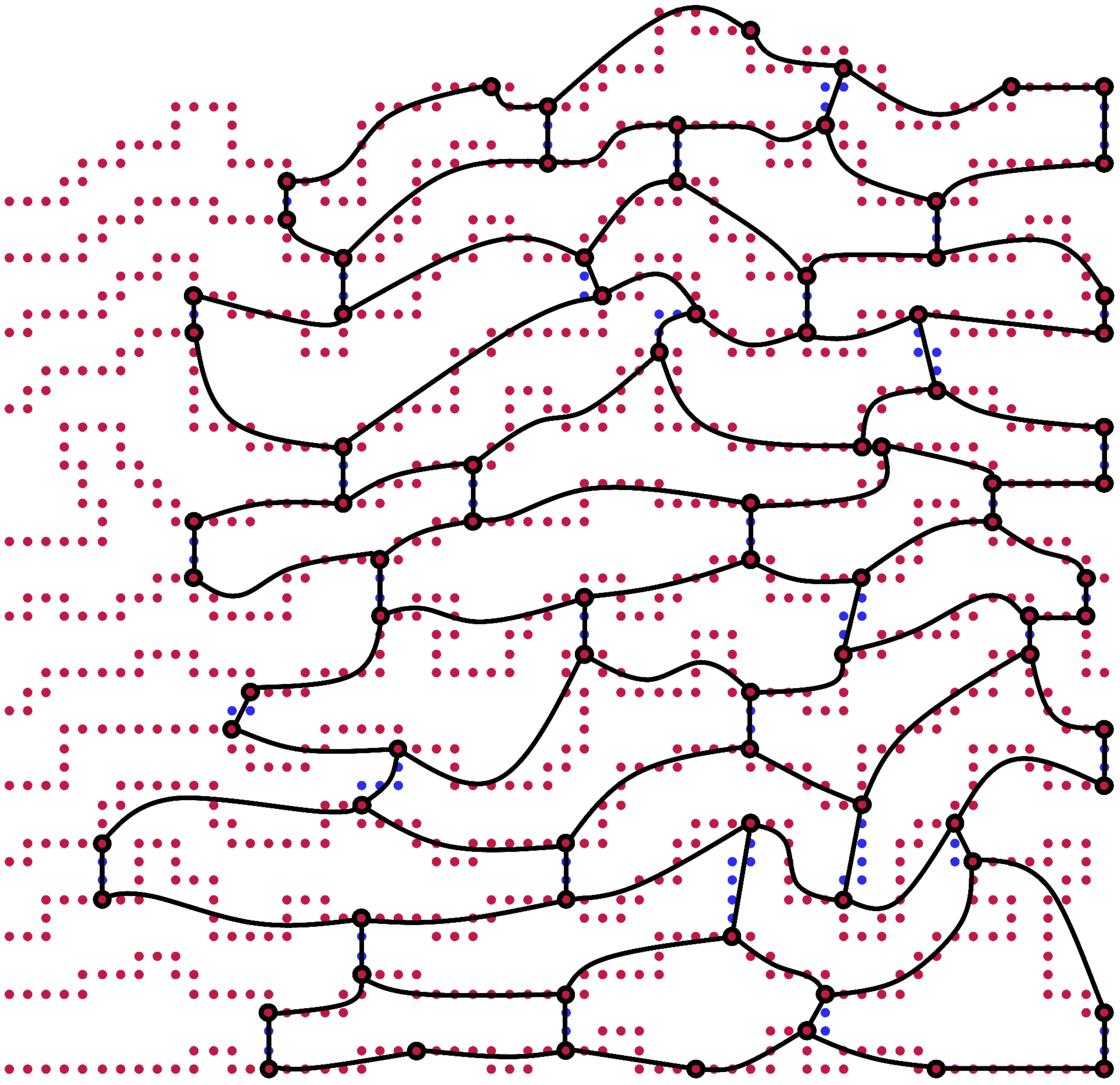}\\
(e) correction of local errors (C.3) &$\quad$&
(f) deletion and contraction (Q.1 \& Q.2)
\end{tabular}
\caption{Summary of the algorithm, illustrated for $p=.85$.  Part (a) shows the initial computational resource: a faulty square lattice in the supercritical phase.  Lost qubits are left blank, while available qubits are yellow.  Part (b) shows the identification of the horizontal (in red) and vertical (in blue) crossings by using a wall-following algorithm, as computed in stage C.1 of the algorithm.  Overlapping crossings are in purple.  Part (c) demonstrates the bridge decomposition, and (d) the alternating bridge decomposition for these crossings, which comprises stage C.2 of the algorithm.  At this point, all of the remaining obstacles to obtaining a perfect hexagonal lattice are localized to the neighborhood where the bridges (in blue) meet the horizontal crossings.  Part (e) corrects the remaining local errors; this is stage C.3.  Finally, part (f) show stages Q.1 and Q.2, where the identified vertices are isolated by performing local quantum measurements on the remaining vertices.  These are either deleted or contracted, until the final state is a hexagonal lattice cluster state.}
\label{F:algorithm}
\end{center}
\end{figure*}

\section{Results}

In this section we first show that in the subcritical phase the faulty cluster is not a universal resource for one-way quantum computation.  Next we overview our algorithm that concentrates a perfect cluster state from a faulty cluster in the supercritical phase and discuss the efficiency of this algorithm, finding that it scales optimally.  Finally, we show that the {\it entanglement width\/}~\cite{Van-den-Nest2006,Van-den-Nest2007a} is an order parameter for this phase transition in computational power.

\subsection{Non-universality and classical simulatability in the subcritical phase}\label{S:subcritical}

In the subcritical phase (i.e., $p < p_c$), the computational power of the resource state is not universal (see the proof by an entanglement criterion in Sec.~\ref{S:phase_transition}), and indeed any one-way quantum computation on it can be simulated efficiently by classical methods. The key observation is that in the subcritical phase, the largest connected component in the graph is almost surely $\cO{\log N}$ and its standard deviation is known to be smaller than $\cO{\log\log N}$ and conjectured to be~$\cO{1}$~\cite{Bazant2000}. Since there are at most $\cO N$ components, the computational subspace has dimension bounded by ${\rm poly}(N)$ and therefore can be simulated classically.

Note that efficiency is a crucial issue here, as the size of the largest connected component still grows unboundedly with $N$, so that with an exponential overhead in the resource size, and an exponentially long time to locate the largest component (from the list of holes), we may still be able to implement any quantum computation. However, we only consider a family of resources to be universal if they can be used to generate the output of any quantum computation with polynomial spatial and temporal overhead.

\subsection{Universality in the supercritical phase}\label{S:universality}

In this section, we describe our algorithm for turning a faulty cluster state in the supercritical phase into a perfect hexagonal cluster state. Additionally, we look at the efficiency of this procedure and show that it achieves the optimal scaling (that is, linear in $N$). We therefore show that in the supercritical phase the faulty cluster state is a universal resource for quantum computation.

\subsubsection{Overview of the algorithm}\label{S:overview}

Our algorithm has two distinct components.  First there is a classical processing stage where we identify a subset of qubits in the faulty square lattice whose topological minor is a regular hexagonal lattice.  Second, we apply a sequence of quantum measurements to concentrate a true hexagonal lattice state from the identified subset.

It is easiest to explain the quantum part of this algorithm first.  Assuming that the classical part of the algorithm has identified a valid subset of qubits, the first step is to simply measure $Z$ on all the qubits outside of this subset, thus disconnecting them from the graph.  Now since what is left has a hexagonal lattice for a topological minor, we can measure $Y$ on all the degree-2 vertices in our remaining graph to topologically contract it into a perfect hexagonal lattice, up to outcome-dependent local unitaries.

To construct the classical algorithm, we recall that although $Z$ measurements allow us to remove vertices, we have no way to remove edges between members of our subset.  Therefore neighboring qubits in the identified subset cannot have edges that break the topological properties of our subgraph.  Because we cannot tolerate such rogue edges, we need to be careful about which qubits we decide to keep in our subset.
With these complications in mind, we subdivide the classical part of the algorithm into three parts: path identification, error localization, and correction of local errors.  In the first step, we identify $\cO L$ disjoint paths across the lattice, both horizontally and vertically, and ensure that they don't interfere with themselves or their neighbors.  Next we ensure that the global topology is that of a hexagonal lattice, modulo local errors that might occur near where horizontal and vertical paths meet.  Finally, we eliminate all the remaining local errors.

In summary, our algorithm proceeds as follows:
\begin{enumerate}
\item[] Lattice Identification (Classical)
\begin{enumerate}
\item[C.1)] path identification
\item[C.2)] error localization
\item[C.3)] correction of local errors
\end{enumerate}
\item[] Lattice Contraction (Quantum)
\begin{enumerate}
\item[Q.1)] deletion (measure $Z$)
\item[Q.2)] contraction (measure $Y$)
\end{enumerate}
\end{enumerate}
Each of these steps is elaborated rigorously in the appendix and illustrated in Figure~\ref{F:algorithm}, but it is worthwhile to give an informal description of the classical part (steps C.1 -- C.3) here.

To identify the $\mathcal{O}(L)$ horizontal and vertical paths in step C.1, we use a right-handed wall follower algorithm.  Intuitively, this is the method of solving a maze where the person walking through the maze keeps their hand on the right-hand wall at all times.  This method finds the right-most (boundary) path in a given faulty lattice, and through successive applications finds a maximal set of disjoint paths which, according to percolation theory, must be size $\mathcal{O}(L)$.  For practical reasons, we use only a constant fraction of these paths, which can be seen in Figure~\ref{F:algorithm}(b).

At this point in the algorithm (C.2) we may have extra edges we don't want in our subgraph.  In order to localize these errors to the regions where horizontal and vertical paths intersect, we start by making what is called a bridge decomposition, as in Figure~\ref{F:algorithm}(c).  Informally, a bridge is a piece of a vertical path that connects adjacent horizontal paths such that it only enters the vicinity of each horizontal path at one point. By throwing away every other bridge we ensure the global topology is that of a hexagonal lattice except for the intersection regions, as seen in Figure~\ref{F:algorithm}(d).

Finally in stage C.3 we correct local errors by taking the shortest horizontal path across the local intersection region; see Figure~\ref{F:algorithm}(e).  It can be shown that this takes care of all remaining superfluous edges.

\subsubsection{Efficiency}

\begin{figure}[t]
\includegraphics[width=8cm,clip]{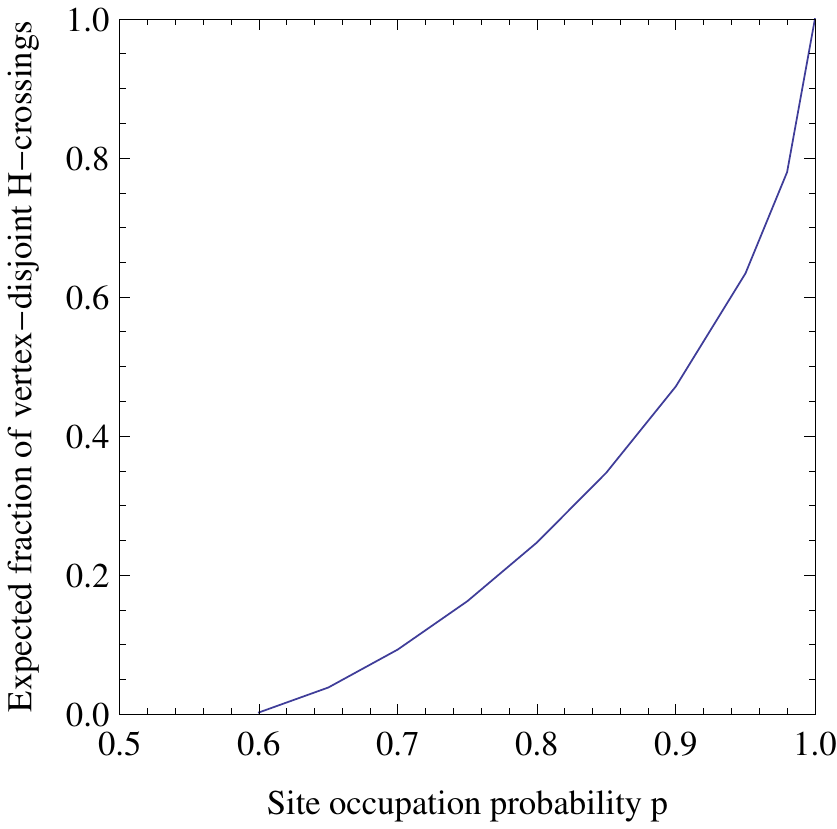}
\caption{Expected number of vertex-disjoint horizontal crossings normalized by the linear size of the lattice $L$ as a function of the site occupation probability $p$.  This is for site percolation on the square lattice in the large $L$ limit.  The ideal expected overhead is therefore the inverse of this number, but squared to account for the vertical crossings as well.  Due to practical concerns in our algorithm, the actual overhead may differ by a small additional constant factor, as discussed in the text.}\label{F:overhead}
\end{figure}

As mentioned in Section~\ref{S:perctheory}, there are almost surely $\cO L$ crossings in both the horizontal and vertical directions.  Stage C.1 of our algorithm identifies {\em all\/} of these crossings, and subsequent stages keep, in the worse case, 1/3 of these.   This factor of 1/3 comes from the necessity of keeping neighboring paths sufficiently far apart that they don't interfere with each other, as discussed in detail in the appendix.  Thus, our final lattice is $\cO L \times \cO L$ in spacial extent, which is obviously optimal up to a constant factor.

The expected size of the final perfect lattice depends on how far above the percolation threshold $p$ is.  This introduces an overhead which is constant with respect to $L$, but increases as $p - p_c$ approaches zero.  We numerically estimated this overhead for all $p$ in Figure~\ref{F:overhead}.

Our algorithm also runs in $\cO N$ time steps; a careful discussion of this scaling is deferred to the appendix.

\subsection{Phase transition in entanglement width}\label{S:phase_transition}

We would like to characterize the transition to universality as a phase transition, since clearly one occurs at the percolation threshold, but we wish to find an order parameter that has an operational meaning in terms of quantum computation. A likely candidate for such a function is an entanglement measure due to the relationship between entanglement and quantum computation.

We utilize the entanglement width $E_{\rm wd}$, introduced in Refs.~\cite{Van-den-Nest2006,Van-den-Nest2007a}, to evaluate the amount of entanglement in a resource state. The entanglement width is an entanglement measure that does not increase
during a one-way quantum computation (more precisely, it is a type II entanglement monotone in the terminology of Ref.~\cite{Van-den-Nest2007a}). Basically, it is the bipartite entanglement entropy maximized over all possible partitions induced by a subcubic tree, and then minimized over all such trees. We can take the bipartite Schmidt rank instead of the bipartite entropy as the base of the entanlgement width, since both are equivalent for graph states.

It was shown in Ref.~\cite{Van-den-Nest2007a} that, in order for a family of $N$-qubit states to be {\it efficiently} universal for one-way quantum computation, it must have an entanglement width that scales faster than polylogarithmic in $N$. Indeed, the entanglement width behaves as  $E_{\rm wd} \geq \cO{\sqrt{N}}$ for the 2D cluster state or the 2D hexagonal lattice state of total size $N$, which are efficiently universal\footnote{We briefly mention another independent proof for this lower bound of the entanglement width of the 2D cluster state, which was obtained in Ref.~\cite{Van-den-Nest2007a} using graph theory. The proof consists of two steps. First, it is readily shown that the so-called planar code state can be obtained by local measurements and classical communication {\it with a constant overhead} from the 2D cluster state. Second, it has been shown in Ref.~\cite{Bravyi2007} that the entanglement width of the planar code state of size $N=L \times L$ is lower bounded by $\sqrt{N}$, by taking advantage of the recent intensive study on the area law of entanglement. Thus, the lower bound of $\cO{\sqrt{N}}$ holds true for the 2D cluster state, since the entanglement width is an entanglement monotone.
This concludes the proof. Note however that this planar code state is an example which illustrates the entanglement criterion is necessary but is not sufficient in general, since measurement-based computation on it was shown to be classically simulatable efficiently \cite{Bravyi2007}.}. Furthermore, for any state of $N$ qubits the entanglement width has the same upper bound as the entanglement entropy, namely $\cO N$.  A key property of the entanglement width relevant to our analysis is that, for a product state of the whole system,
\begin{equation}\label{eq:Ewd_max}
	E_{\rm wd}\bigg(\bigotimes_j |\psi_j\rangle\bigg) =
	\max_j \; E_{\rm wd}\big(|\psi_j\rangle\big).
\end{equation}
In other words, we need only consider the most entangled connected component to evaluate the total entanlgement.

In the subcritical phase, as noted in Sec.~\ref{S:subcritical}, the largest connected component is of size $\cO{\log N}$. Therefore the entanglement width for this component cannot scale greater than $\cO{\log N}$, and hence this upper bounds the scaling for the entire faulty lattice. On the other hand, in the supercritical phase, our concentration algorithm essentially performs a one-way computation that produces a perfect 2D cluster state of size $\cO N$, whose entanglement width is lower bounded by $\cO{\sqrt N}$. Since the entanglement width is an entanglement monotone which
cannot increase during a one-way quantum computation, we have shown that
the entanglement width of the original faulty resource state is lower bounded
by $\cO{\sqrt N}$, as well. Namely, the constant overhead of our algorithm does not reduce the scaling of the lower bound.

The fact that in the subcritical phase $E_{\rm wd} \leq \cO{\log{N}}$ and in the supercritical phase $E_{\rm wd} \geq \cO{\sqrt{N}}$ indicates that the entanglement width is a proper continuous function that can be taken as an order parameter for the phase transition to universality. The amount of entanglement relevant to the computational power of the resource state drops exponentially at the threshold, and actually below the threshold any measurement-based computation on the resource is found to be simulatable efficiently by classical computers as seen in Section~\ref{S:subcritical}.

This phenomenon is surprising, since one might expect the amount of entanglement, or the value of the resource state for computation, degrades {\it gradually} according to the imperfection. In this respect, our observation is reminiscent of a classic paper by Aharonov \cite{Aharonov2000} on a quantum-to-classical transition in a mixed quantum computer in which an ``entanglement length'' was shown to change from being infinite to finite at not lower than some percolation threshold (see also \cite{Raussendorf2005}). The quantum-to-classical transition in Refs.~\cite{Aharonov2000,Raussendorf2005} can be seen as a phase transition {\em at a finite temperature}, in which the noise destroys quantum coherence making the quantum state highly mixed and therefore useless for quantum computation.

By contrast, our work views the phase transition in light of an operational order parameter which is motivated by quantum computing~\cite{Van-den-Nest2006,Van-den-Nest2007a}, and we consider only pure states.  The infinite entanglement-length condition employed in the previous works does not necessarily guarantee the capability of universal quantum computation, despite guaranteeing non-classical states.  While our order parameter is also insufficient for guaranteeing universality in general, it has a clear operational meaning, and our algorithm enables us to prove universality explicitly in our model, thanks to its optimal resource scaling.

We note that in order to observe a phase transition in the computational power of the resource states, the selection of a proper entanglement measure was crucial. One might wonder whether other kinds of continuous entanglement measures that are useful in forming necessary criteria for universal resource states can also be used as order parameters.  For example, the geometric measure, among the type II measures considered in Ref.~\cite{Van-den-Nest2007a}, is a multipartite entanglement measure widely studied in the literature. It gives the maximum value $\lfloor \frac{N}{2}\rfloor$ for the 2D cluster state of $N$ qubits, as shown in \cite{Markham2007}. However, this measure is shown to be extensive for the product of graph states; namely $E(\bigotimes_j |G_j\rangle) = \sum_j E(|G_j\rangle)$, so that the entanglement in small connected components contributes to the total entanglement. Thus we do not see a drastic change in this measure in the vicinity of the threshold. This is similar to the way the non-universality of the 1D cluster state of $N$ qubits  is detected by the entanglement width but not the geometric measure~\cite{Van-den-Nest2006}.

\section{Conclusion}

Starting from a faulty $L \times L$ cluster state which contains heralded hole defects with probability $1-p$, we showed how to concentrate a perfect $\cO L \times \cO L$ cluster state in the supercritical percolation phase using only a polynomial amount of preprocessing.  We also showed that quantum computation in the subcritical phase admits an efficient classical simulation.
We gave an interpretation of these results by using the entanglement width as an order parameter for the phase transition to universality, occurring at the percolation threshold when the amount of entanglement relevant to the computational power changes exponentially for the system size.

It would be very interesting to consider generalizing our algorithm to include other lattice geometries, in both 2 and 3 dimensions, as well as bond percolation models.  While extensions to other 2D geometries seem possible, extensions to 3D might require substantial new ideas because our proof relies heavily on the planarity of the underlying lattice.

In closing, we offer the following quote as a poetic description of our algorithm for the supercritical phase.

{\it ``In every block of marble I see a statue as plain as though it stood before me, shaped and perfect in attitude and action. I have only to hew away the rough walls that imprison the lovely apparition to reveal it to the other eyes as mine see it.''\/}
--Michelangelo

\acknowledgments

DEB, STF, AM, and AJS would like to thank H.~Briegel, W.~D\"ur, and M. Van den Nest for their hospitality and helpful discussions at the University of Innsbruck, where this work was initiated.  MBE and STF were supported by ARO Contract No.~W911NF-04-1-0242 and NSF Grant No.~PHY-0653596, STM was supported by DTO No.~DAAD19-13-R-0011, AM was supported by the Austrian Science Foundation (FWF), the European Union (OLAQUI, SCALA, QICS), the Austrian Academy of Sciences (\"{O}AW), and AJS worked on this paper whilst employed by the University of Bristol. Both DEB and AJS were supported by the UK EPSRC `QIP IRC' programme. AJS would also like to acknowledge support from a Royal Society University Research Fellowship, and DEB and AM would like to acknowledge funding from the EPSRC Network on Semantics of Quantum Computation (EP/E00623X/1).

\appendix*

\section{Detailed proof of universality}\label{S:details}

In this appendix we give a detailed proof that the algorithm outlined in Sec.~\ref{S:overview} produces a hexagonal lattice from an initial faulty square lattice, and show that it does so efficiently.

\subsection{Definitions}\label{S:definitions}

An {\it H-path\/} is a path from the left-hand side of the lattice to the right-hand side, and similarly for a {\it V-path\/}.  In our algorithm the paths we identify are disjoint, so we can label them successively from bottom to top (for H-paths) or from left to right (for V-paths).  We denote the $j^{\rm th}$ H-path by $H^j$ and the $k^{\rm th}$ V-path by $V^k$.  Note that we use the letters $j$ and $k$ for indexing horizontal and vertical paths respectively.

Given a set of non-intersecting paths, we call them {\it valid\/} if all edges in the original faulty square lattice that connect vertices in the paths are included as edges in the paths.
Less formally, valid paths are those that don't get ``too close'' to themselves or each other.

A valid H-path $H^j$ succumbs to a natural ordering given by $H_m^j > H_{m'}^j$ if $m>m'$, where $H_m^j$ denotes the $m^{\rm th}$ vertex in the path. Similar statements hold for V-paths. From this ordering we get a partial ordering for sets of vertices in $H^j$ by $A \geq B$ if either $H_a^j > H_b^j$ for all $a \in A$ and $b \in B$ or the sets $A$ and $B$ are equal.

A {\it $j$-stripe\/} is the region of vertices lying strictly between $H^j$ and $H^{j+1}$.
The {\it $k^{\rm th}$ $j$-bridge\/}, written $B^{j,k}$, is the segment of the $k^{\rm th}$ vertical path which crosses the $j$-stripe in the following manner. Among the ordered elements of $V^k$, choose the last element that lies in the neighborhood of $H^j$.  This is called the start of the bridge, $s$.  Similarly, choose the first element after $s$ to reach the neighborhood of $H^{j+1}$ and call this the end of the bridge $e$. Then $B^{j,k}$ is the sub-path of $V^k$ with elements lying between $s$ and $e$.  The {\it bridge decomposition\/} $B^k$ for a vertical path $V^k$ is the ordered set of each of the bridges across all the stripes; i.e.
\begin{align}
	B^k = \{B^{1,k},B^{2,k}, \ldots , B^{J-1,k}\}.
\end{align}
Note that $B^k$ is \emph{not} a crossing path, since it is not connected. The {\it complete bridge decomposition\/} is simply the union of the bridge decompositions for each V-path.

The \textit{$(j,k)$-abutment\/} is the part of $H^j$ that neighbors the bridges $B^k$ and is defined as
\begin{align}
	A^{j,k} = H^j \cap \mathcal{N}(B^k).
\end{align}
We also have {\it upper\/} and {\it lower\/} abutments, ${^{\uparrow}\!}A^{j,k}$ and ${^{\downarrow}\!}A^{j,k}$, defined as,
\begin{align}
	{^{\uparrow}\!}A^{j,k} &= H^j \cap \mathcal{N}(B^{j,k}) \\
	{^{\downarrow}\!}A^{j,k} &= H^j \cap \mathcal{N}(B^{j-1,k}),
\end{align}
so that $A^{j,k}= {^{\uparrow}\!}A^{j,k} \cup {^{\downarrow}\!}A^{j,k}$.

\begin{figure}[t!]
\begin{center}
\begin{tabular}{c}
\includegraphics[width=8cm]{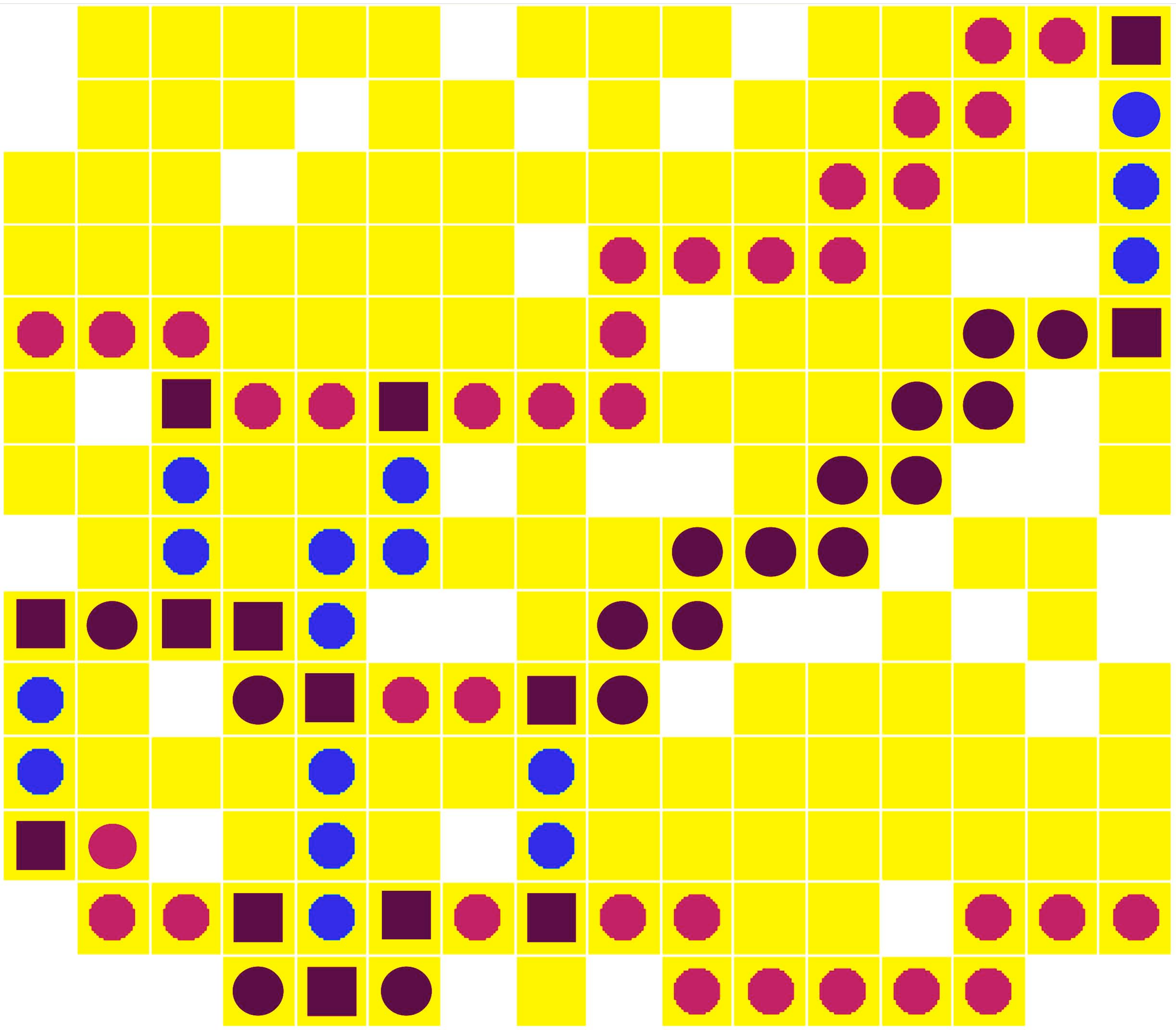}\\
\includegraphics[width=7cm]{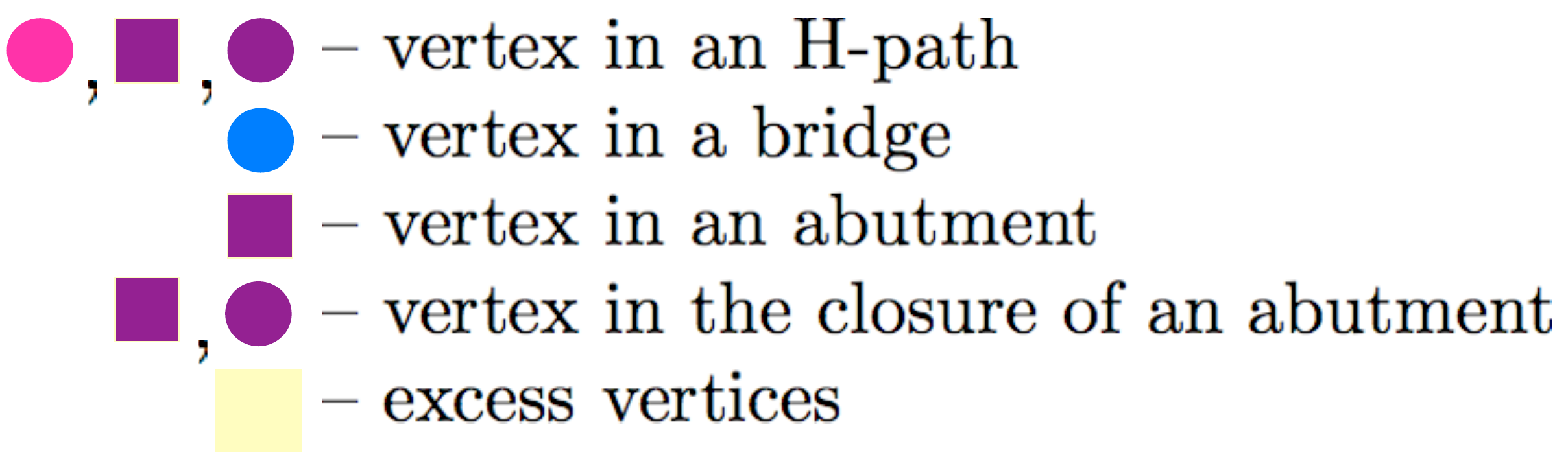}
\end{tabular}
\caption{Illustration of the basic definitions from Section \ref{S:definitions}.}
\label{F:definitions}
\end{center}
\end{figure}

The {\it closure\/} of an abutment $A^{j,k}$ is the set of all vertices $H^j_{m}$ such that there exists vertices $H^j_{m_1},H^j_{m_2} \in A^{j,k}$ such that $m_1 \leq m \leq m_2$.  Intuitively, this is all those vertices lying on the path $H^j$ between the beginning and end of the abutment. Similarly, we can define the closure of just the upper or lower abutment. All of the concepts introduced so far in this section are illustrated in Figure~\ref{F:definitions}.

The strategy we use for finding H- and V-paths is to use the wall following technique familiar from maze solving.  We define the algorithm {\it right-hand wall follower\/} (RHWF) to be the algorithm that finds crossings by traversing a (planar) graph around its boundary counterclockwise.  We can also define a version that follows extremal paths outside the $k$ neighborhood of the boundary. We call this the $k$-local RHWF.  In the algorithm, we use 2-local RHWF to ensure that the neighborhoods of adjacent H-paths do not intersect.

\subsection{Error taxonomy}

There are three types of errors: {\it Closeness\/} errors, {\it degree\/} errors and {\it lattice\/} errors.  A closeness error occurs when a path gets too close to itself or another path.  A degree error is where a vertex is not degree three, as is proper for a hexagonal lattice.  A lattice error happens when the global topology is not that of a regular hexagonal lattice.  Additionally, the first two error classes are subdivided according to what type of paths or degrees are involved, as explained below.

Closeness errors are subdivided as follows.
\begin{itemize}
\item {\it Self-H\/} errors, where a single horizontal path gets too close to itself.  Similarly, there are {\it Self-V\/} errors.
\item {\it H-H\/} errors, where a horizontal path gets too close to another disjoint horizontal path.  Similarly, there are {\it V-V\/} errors.
\item {\it H-V\/} errors, where a horizontal path gets too close to a vertical path, except for a single intersection.
\end{itemize}

Note that valid H-paths have no Self-H or H-H errors (and similar for V-paths).

Degree errors are also subdivided.  At the end of the entire algorithm, we want every vertex to be degree 3 (except on the boundary of the lattice, which is dealt with separately).  Thus, we need to eliminate vertices of degrees 1,2 and 4. {\it Degree 1\/} errors correspond to ``dangling'' qubits while {\it Degree 2\/} errors correspond to ``wires'', both of which can be contracted by measuring~$Y$.  While these are handled in the quantum stage of the algorithm, {\it Degree 4\/} errors are handled in the classical stage.

\subsection{Proof for the classical part}\label{S:proof}

As discussed in Section~\ref{S:overview}, the classical part of our algorithm proceeds in three main stages: path identification, error localization, and correction of local errors. After this is complete, we will have identified a subset of the qubits whose topology is that of a hexagonal lattice, and the quantum part of the algorithm (discussed in the next subsection) will contract this lattice into a hexagonal lattice cluster state.

\begin{figure}[t!]
\begin{center}
\includegraphics[width=8cm]{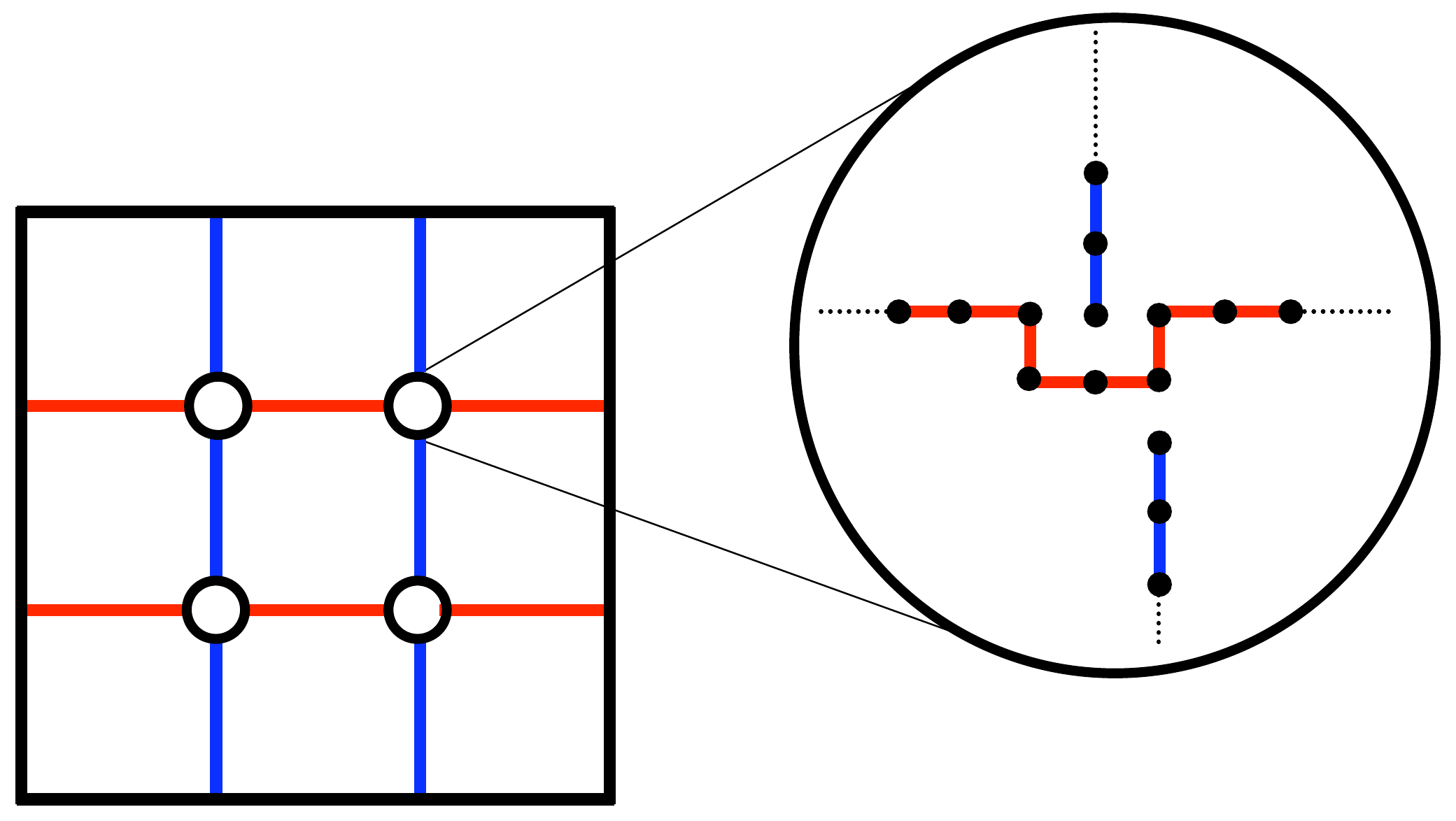}
\caption{Localization of errors, as in part C.2 of the algorithm.  By choosing a bridge decomposition, all the closeness errors are localized to the regions near where the vertical and horizontal paths cross.  As shown in the text, except for those regions, the global topology is that of a regular square lattice.}
\label{F:abutments-ordered}
\end{center}
\end{figure}

As discussed in subsection~\ref{S:vertexdisjoint}, we know that there exists $\cO L$ vertex-disjoint H- and V-paths. The wall follower algorithm defined above will necessarily find all these paths since it finds boundary paths \cite{Nishizeki1988}. For the H-paths, we use the 2-local version of RHWF and for the V-paths, we use RHWF, but keep only every third path\footnote{It may seem unusual at this stage that we don't treat the H-paths and V-paths symmetrically.  The reasons for this will become clear later in the proof.}.  Clearly then we do not have H-H or V-V errors. To eliminate self errors we simply find the shortest path amongst the original vertices and their accompanying internal edge sets, which can only contain degree $1$ and $2$ vertices and hence no self errors. Taking the shortest path also cannot introduce any closeness errors because we do not add any vertices, we only delete them. Thus we have shown that after stage C.1 of the algorithm we have identified $\cO L$ valid H- and V-paths.

\begin{figure}[t!]
\begin{center}
\includegraphics[width=8cm]{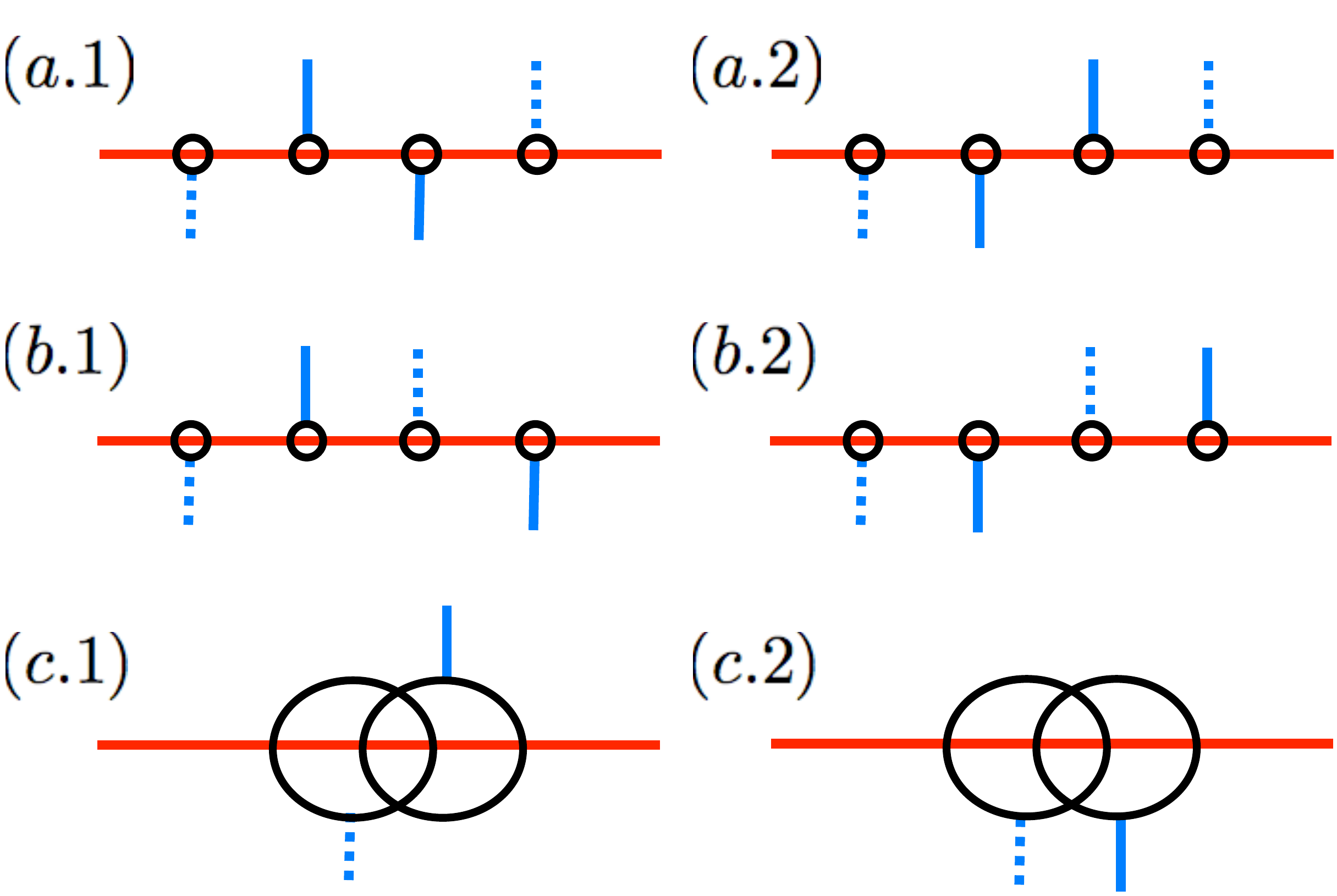}
\caption{Schematic diagram showing the possible ways that the closures of abutments might intersect. The red line is the horizontal path $H^j$, while the blue lines are the bridges incident on $H^j$ for two vertical paths, $V^k$ and $V^{k'}$.  The solid blue line is associated with the index $k$, while the dashed line is associated with $k'$.  Closures of upper and lower abutments are denoted by black circles. In (a) and (b), we assume the closures of the upper and lower abutments are disjoint, but that the total abutments overlap.  In case (c), we treat the possibility that the upper or lower abutment closures themselves overlap.  Note that, while the case of complete containment is not illustrated as a sub-case of (c), it is still treated in the text.}
\label{F:total-order}
\end{center}
\end{figure}

Once this is done, we can find the complete bridge decomposition as defined above. Since we do this by removing vertices while preserving connectivity, this process clearly does not introduce any new closeness errors. This step also localizes H-V errors to the region near the abutments as in Fig.~\ref{F:abutments-ordered}.

The next step is to take an {\it alternating bridge decomposition\/}, which is defined as follows.  In a $j$-stripe we can delete the bridges $B^{j,k}$ such that $j+k$ has odd parity, which is alternating in the sense that we take alternating bridges as indexed by $j$ and $k$. We want to show that the alternating bridge decomposition is also alternating in a stronger, topological sense. We do this by showing below that the natural partial ordering on the abutments is actually a total ordering; that is, each pair of abutments along an H-path can be compared. Intuitively, then, the global topology of the lattice is that of a square lattice except for possible errors localized to the abutments. Taking alternating bridges will then clearly give a hexagonal lattice. This intuition is illustated in Fig.~\ref{F:abutments-ordered}.

Showing that abutments are totally ordered (as sets) follows from showing that closures of abutments are disjoint. We show that they are disjoint by considering the various ways they might overlap and showing that each cannot occur. This leaves several configurations, as illustrated in Fig.~\ref{F:total-order}.

\begin{figure}[t!]
\begin{center}
\includegraphics[width=8cm]{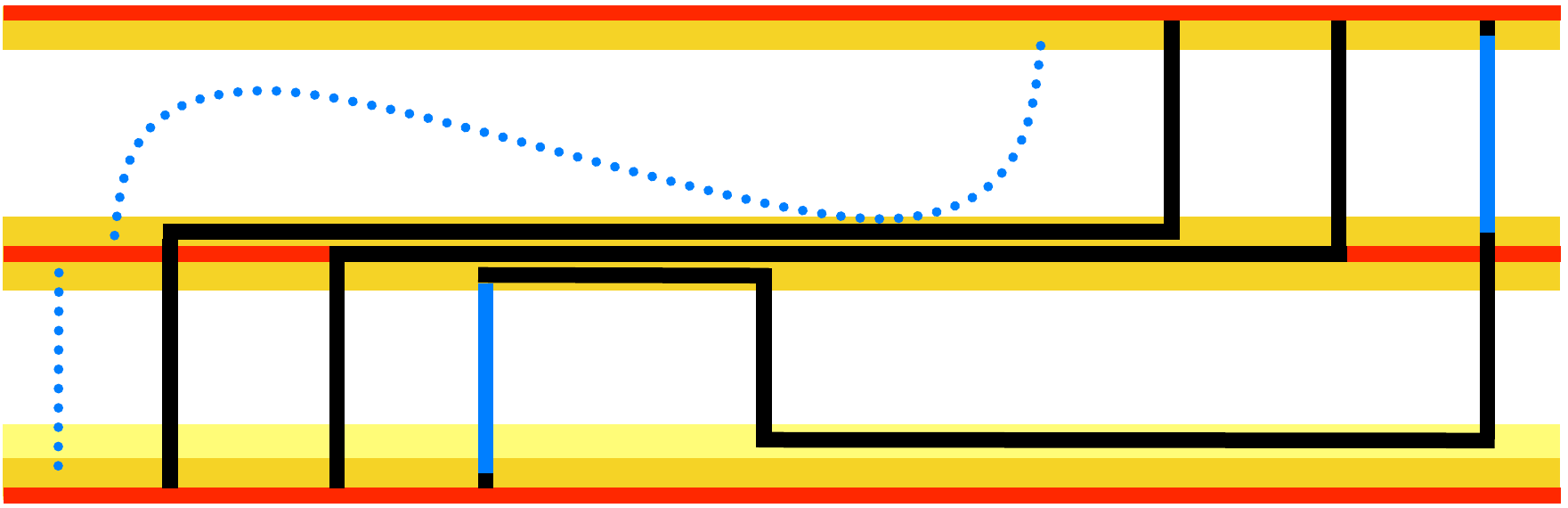}
\caption{Detailed examination of case (b.2) from Figure~\ref{F:total-order}. Horizontal paths are colored red, and their 1- and 2-neighborhoods are colored different shades of yellow. (We only highlight neighborhoods that are used in the proof.) Vertical paths are denoted by black and/or blue. Stage C.1 of the algorithm keeps only the vertical path with the blue bridges, and stage C.2 keeps only the bridges from this path. There is no room for the dashed line to create a bridge in the upper stripe that would result in an error of the form depicted in Figure~\ref{F:total-order}(b.2).}
\label{F:bridges-alternate}
\end{center}
\end{figure}

The cases in Fig.~\ref{F:total-order}(a.1), (a.2), and (b.1) can only occur if two V-paths cross, but this does not happen because there are no V-V errors left after stage C.1.

The only possible configuration for Fig.~\ref{F:total-order}(b.2) is more carefully illustrated in Fig.~\ref{F:bridges-alternate}.  The fact that this is the only possible configuration follows from first considering Figure~\ref{F:total-order}(b.2) and drawing in the two V-paths that were deleted in stage C.1. Because the configuration in Figure~\ref{F:total-order}(a.2) can't happen, we must draw Figure~\ref{F:bridges-alternate} so that the bridges connect as shown. The V-path furthest to the right cannot go any lower without creating a new, further right, bridge.  The next furthest V-path must lie along the middle H-path, because if it went any lower, then RHWF would have found that lower path when identifying the middle H-path. (This isn't an issue for the furthest right V-path because it enters the 2-neighborhood of the lowest H-path.) The last existing V-path cannot intersect the others, and so must remain at least as high as shown in Figure~\ref{F:bridges-alternate}.  Therefore this configuration is exhaustive.

We now show that the configuration in Figure~\ref{F:bridges-alternate} leads to a contradiction, and so case (b.2) cannot occur.  Since in stage C.1 of the algorithm, we take every third path, the two black vertical paths necessarily exist and don't contribute bridges.  To obtain a diagram like that in Fig.~\ref{F:total-order}(b.2), we would need a fourth vertical path, the assumed dashed blue line, that has the following property: its bridge in the lower stripe must lie to left of the lower blue bridge, while its bridge in the upper stripe must lie to the right of that same lower blue bridge. To form a bridge in the appropriate place in the upper stripe, this assumed fourth path would need to pass through the neighborhood of the middle H-path.  But since V-paths aren't allowed to cross, this neighborhood is blocked by one of the deleted (black) V-paths.  Therefore, case (b.2) cannot occur.

\begin{figure}[t!]
\begin{center}
\includegraphics[width=8cm]{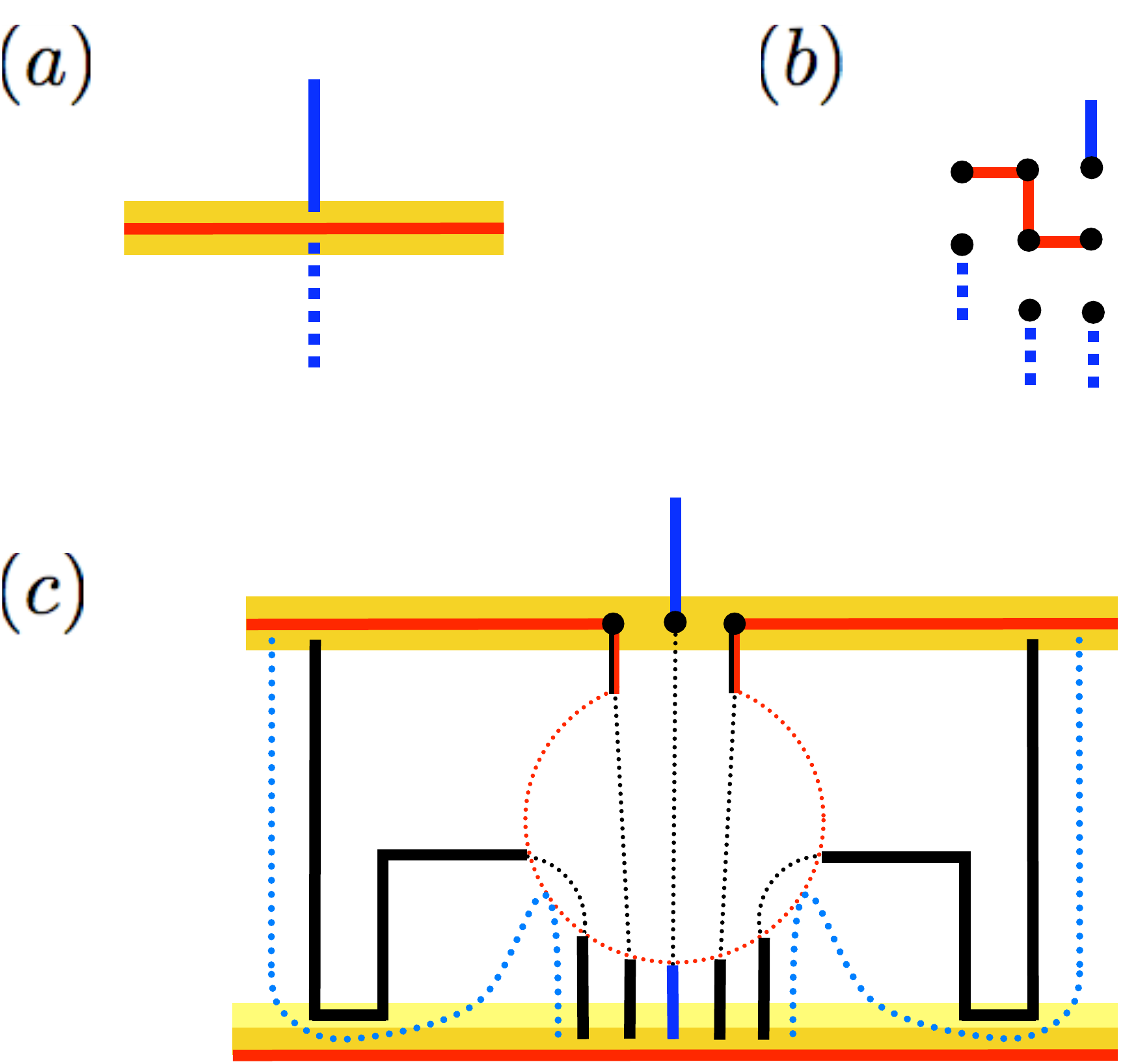}
\caption{Illustration of the configurations that lead to Figure~\ref{F:total-order}(c.1).  In configuration (a), where $\ell=1$, there is a clear violation of the requirement that we have kept every third path.  In (b), where $\ell=3$, regardless of which dashed path is chosen there is again not enough room for two more V-paths. Finally, in case (c), which covers all $\ell>3$, we have the dreaded octopus configuration. }
\label{F:three-cases}
\end{center}
\end{figure}

Figure~\ref{F:total-order}(c.1) breaks into the three non-trivial cases in Figure~\ref{F:three-cases}. They are enumerated by the length $\ell$ of the closure of the lower abutment for the solid blue bridge. For case \ref{F:three-cases}(a), where $\ell=1$, there is a clear contradiction with the requirement of keeping every third V-path.  Likewise for case \ref{F:three-cases}(b), where $\ell=3$.  For case \ref{F:three-cases}(c), which cover all cases with $\ell>3$, it becomes slightly more complicated.  The argument that this can't happen is very similar to that discussed above for Figure~\ref{F:total-order}(b.2).  The essential point is that keeping every third V-path doesn't leave enough space for this kind of error.  Thus, the case depicted in Figure~\ref{F:total-order}(c.1) doesn't happen either.

Finally, we have case (c.2).  Since we have kept only every third vertical path, neighborhoods of identified V-paths cannot overlap.  For case (c.2) to occur, the horizontal path would have to enter the neighborhood of the dashed blue bridge, {\it then leave it\/} before entering the neighborhood of the solid blue bridge.  But in order to create an error like (c.2), it would have to {\it re-enter\/} the neighborhood of the dashed blue bridge.  This is a contradiction, since then the H-path would either have to cross itself (eliminated in stage C.1) or a bridge (not possible by the definition of a bridge).

Thus we conclude that the closures of the abutments are indeed disjoint and taking alternating bridges gives a hexagonal lattice structure, up to errors localized to the abutments. In other words, the graph minor achieved by collapsing the region near the abutment has a topological minor which is exactly a regular hexagonal lattice.  Thus, if we can transform these regions to have exactly the topology of a single local vertex in a hexagonal lattice, then the quantum part of the algorithm will succeed.

Now the only step left for the classical stage is to fix the local H-V errors in the region of the abutments. To do this, we can actually enumerate the possibilities. First observe that all vertices in an abutment are degree $3$, other vertices in an H-path are degree $2$, and the endpoint (or starting point) of a bridge, $e$ (or $s$) is either degree $2$, $3$, or $4$.  There is no lattice error if there is one vertex in the region near an abutment of degree $3$ and the rest are degree $2$. Therefore if $e$ has degree $2$ there is no problem; the only degree 3 vertex in this local region is the single vertex belonging to the abutment.

In the other cases, where $e$ has degree 3 or 4, we can eliminate the errors in the following manner.  First delete everything in the H-path that lies in the closure of the abutment \emph{except} for the endpoints of the abutment.  Then, by adding the endpoint of the bridge to the remaining vertices, we once again have a valid H-path, which additionally has no H-V errors, since the only degree 3 vertex is now the endpoint of the bridge, which is now considered as part of the H-path.

Although we have done this procedure for the endpoints, we can also do it for the starting points.  Both of these points can be done separately, since bridges always have distinct starting and ending vertices as a consequence of using the 2-local RHWF.  One final subtle point is that, when the reconfiguration is done for both the start and end of the bridge, this might result in an H-H error.  Because of the restricted nature of these H-H errors, these errors do not affect the global topology; in fact, an example of this subtlety is present in Fig.~\ref{F:algorithm}.

\subsection{The quantum part of the algorithm}

The quantum part of the algorithm is straightforward compared to the classical part.  As stated before, we simply measure $Z$ on vertices that were not identified in the classical part, thus disconnecting them.  Then we measure $Y$ on the remaining degree 1 or 2 vertices, which will topologically contract the lattice.  The only delicate part is the boundary, where we should leave one additional vertex present between each bridge in the H-paths, so that we have a true hexagonal lattice.  A valid vertex is always available because we take every third V-path.

We note that these measurements must be performed sequentially, together with conditional single qubit rotations on the neighboring qubits in order to preserve the graph state~\cite{Hein2004}.

\subsection{The existance of ${\mathcal O}(L)$ vertex-disjoint crossings\label{S:vertexdisjoint}}

We argue that we can expect ${\mathcal O}(L)$ vertex-disjoint crossings in both the horizontal and vertical directions. Let us consider the event (increasing with respect to $p$) that there exists a path between the one side and the opposite side in the 2D square grid. Denote by $m_L$ the maximal number of vertex-disjoint paths crossing the 2D square grid of size $N=L^2$ from one side to the other.

According to a basic result in independent percolation, in the supercritical phase $p > p_c$, the event such that at least a single crossing path exists happens with probability $P_p (m_L \geq 1) \geq 1 - e^{-\alpha(p) L}$, for some $\alpha(p) > 0$, that is, almost surely in the thermodynamic limit.

The existence of a crossing is stable with respect to changes in any $r$ or fewer sites if there exists at least $r+1$ vertex-disjoint crossings.  In order to show that the stability is $r = {\mathcal O}(L)$ with probability close to one in the thermodynamic limit, we utilize an argument for the stability of an event from Section~2.6 of Ref.~\cite{Grimmett1989}, in which the following inequality is proved in the bond percolation setting, but is easily adapted to site percolation, as discussed in Section~5 of Ref.~\cite{Grimmett1990}. We first write $r = \beta L$ with some spatial overhead $\beta > 0$.  Then the existence of $r+1$ vertex-disjoint crossings is still expected with probability close to one given by
\begin{equation}
P_p (m_L \geq \beta L) \geq 1 -e^{-\gamma_\epsilon(p) L},
\end{equation}
as long as the reduced exponent
\begin{equation}
	\gamma_\epsilon(p) = \alpha(p_c+\epsilon) - \beta \log \left(\frac{p}{p-p_c-\epsilon}\right)
\end{equation}
is positive and $p-p_c > \epsilon>0$.  Namely, for a given $p$ in the supercritical phase, we can almost surely expect $m_L$ is ${\mathcal O}(L)$ with a spatial overhead depending only on $p$, since $\beta$ can be chosen optimally over the range of $\epsilon$.

\subsection{Details of efficiency}

In this section we show that each step in the algorithm runs in time at most $\cO N$. First, RHWF runs $\cO{L}$ times to find $\cO{L}$ paths. Each time it finds a path, it will visit any given vertex at most $4$ times, since this is the maximum degree of a vertex.  Furthermore, each time RHWF is called, it can only visit vertices that are distinct from those visited by any previous call.  Since the most number of vertices possible is $N$, all of the RHWF calls will visit no more than $4N$ vertices, which is $\cO N$.  The bridge and alternating bridge decompositions again visit at most $\cO N$ vertices, this time only once, since there is no backtracking.  Reconfiguring the regions near the abutments requires only $\cO N$ steps as well, since it only checks the degrees of the bridge start and end points.  Thus, the entire classical part of the algorithm has run time $\cO N$.  The quantum part only requires $\cO N$ measurements, so the $\cO N$ scaling holds for the entire algorithm.


\end{document}